\documentclass[fleqn,usenatbib]{mnras}

\usepackage{newtxtext,newtxmath}
\usepackage[flushleft]{threeparttable}
 
\usepackage[T1]{fontenc}

\DeclareRobustCommand{\VAN}[3]{#2}
\let\VANthebibliography\thebibliography
\def\thebibliography{\DeclareRobustCommand{\VAN}[3]{##3}\VANthebibliography}

\usepackage{graphicx}	
\usepackage{amsmath}	
\usepackage{anyfontsize}
\usepackage{stackengine}
\usepackage{threeparttable}
\usepackage{xcolor}

\usepackage[normalem]{ulem}
\usepackage{array}
\newcolumntype{L}[1]{>{\raggedright\let\newline\\\arraybackslash\hspace{0pt}}m{#1}}

\newcommand{\ergs}{\mbox{ erg s}^{-1}}

\title[Winds and jets in magnetized media]{A comparative study of radio signatures from winds and jets: Modelling synchrotron emission and polarization}

\author[Meenakshi et. al]{Moun Meenakshi$^{1}$,\thanks{E-mail:mounmeenakshi@iucaa.in (MM)}
Dipanjan Mukherjee$^{1}$,\thanks{E-mail:dipanjan@iucaa.in (DM)}, Gianluigi Bodo$^{2}$, Paola Rossi$^{2}$ and Chris M. Harrison$^3$
\\
$^{1}$Inter-University Centre for Astronomy and Astrophysics, Pune- 411007, India\\
$^{2}$INAF, Osservatorio Astroﬁsico di Torino, Strada Osservatorio 20, I-10025 Pino Torinese, Italy\\
$^{3}$School of Mathematics, Statistics and Physics, Newcastle University, Newcastle upon Tyne NE1 7RU, UK
}

\date{Accepted XXX. Received YYY; in original form ZZZ}

\pubyear{2024}

\begin{document}
\label{firstpage}
\pagerange{\pageref{firstpage}--\pageref{lastpage}}
\maketitle

\begin{abstract}

Outflows driven by active galactic nuclei (AGN) are seen in numerous compact sources; however, it has remained unclear how to distinguish between the driving mechanisms, such as winds and jets. Therefore, our study aims to offer observational insights from simulations to aid in this distinction. Specifically, in this paper, we investigate the evolution of wide-angled, moderately relativistic, magnetized winds and analyze their non-thermal radio emission and polarization properties. We find that the evolution of winds varies depending on factors such as power, density, and opening angle, which in turn influence their observable characteristics. Additionally, different viewing angles can lead to varying observations. Furthermore, we note distinctions in the evolution of winds compared to jets, resulting in disparities in their observable features. Jets typically exhibit a thin spine and hotspot(s). Winds manifest broader spines or an ``hourglass-shaped'' bright emission in the cocoon, which are capped by bright arcs. Both display high polarization coinciding with the bright spine and hotspots/arcs, although these regions are relatively compact and localized in jets when compared to winds. We emphasize the importance of high resolution, as we demonstrate that emission features from both jets and winds can become indistinguishable at lower resolutions. The distribution of polarization is largely unaffected by resolution, though lower polarization becomes more noticeable when the resolution is decreased.

\end{abstract}

\begin{keywords}
galaxies: active - ISM: jets and outflows - MHD - turbulence - methods: numerical
\end{keywords}



\section{Introduction}
Outflows driven by the Active Galactic Nuclei (AGN) have been observed through multi-wavelength studies in a plethora of sources \citep{tombesi_2015,nesvadba_2021,zovaro_2019,vayner_2021,girdhar_2022,shen_2023}. These outflows can arise due to either collimated jets or wide-angled winds resulting from the activity of the central supermassive black hole \citep{harrison_2024}. Theoretical models and observations indicate that both jets and winds are launched by distinct mechanisms. The Blandford-Znajek mechanism is attributed to the generation of jets from the CSMBH \citep{blandford_1977}, rendering them collimated and highly magnetized. The formation of disc winds is explained by three popular models, which are radiation driving \citep{murray_1995,proga_2000}, magnetic driving \citep{blandford_1982,yuan_2015,wang_2022} and thermal driving \citep{begelman_1983}. These winds are wide-angled outflows, as presented in the models from \citet{faucher_2012} and \citet{nims_2015}.

The primary source of emission from both winds and jets is attributed to shocks that accelerate non-thermal electrons. These accelerated electrons generate incoherent synchrotron radiation, which is detected as radio emission \citep{blandford_1978,melrose_1993,mukherjee_2020,mukherjee_2021}. 
A recent survey by \citet{sabater_2019} has shown that almost all high mass galaxies ($M > 10^{11} M_\odot$) show radio emission beyond a threshold limit of $10^{21} \mathrm{W\, Hz^{-1}}$. In radio-loud sources\footnote{``radio-loud'' sources are characterized by high radio luminosity (e.g. $P_{1.4~\mathrm{GHz}}>>10^{24}~\mathrm{W~Hz^{-1}}$) and high ratio-to-optical flux density ratio (>10), and vice-versa for radio-quiet sources \citep{padovani_2017}.}, the jets are expected to dominate the radio emission, whereas the origin in radio-quiet AGN is widely debated \citep{panessa_2019,kharb_2023}. In such sources, it can depend on various mechanisms, including synchrotron emission from electrons accelerated in low-power magnetized jets \citep{falcke_2000,gallimore_2006}, wide-angle winds \citep{nadia_2016,peng_2020}, and star formation \citep{vries_2007,condon_2013}.

Polarimetry studies \citep{sebastian_2020} have found distinctions in polarization features that enable differentiation between radio emissions arising from star formation and AGN-driven outflows. Several recent studies indicate that traditional definitions of radio-quiet may not always mean a lack of AGN winds or jets \citep[see e.g.][]{jarvis_2019,jarvis_2021,silpa_2020,silpa_2023}; nonetheless, in several instances, the nature of the emission remains ambiguous. A potential solution lies in the radio emission and polarization study of resolved jets/winds, as it can aid in distinguishing between various radio emission mechanisms. The expectation stems from the understanding that jets and winds are launched through different mechanisms and are therefore expected to interact differently with their surroundings. Consequently, when these two outflows encounter a similar ambient medium, observable differences in their emission and polarization characteristics may arise, although this area remains largely unexplored. This serves as the motivation for our study (also discussed in \citet{meenakshi_2023}, hereafter referred to as Paper I) of jets and winds spanning a few kpc in scale, where we aim to analyze their observable properties.

In Paper I, we focused on magnetized jets interacting with turbulent magnetic fields. The effect of magnetized Shocked Ambient Medium (SAM) on the net polarization from the jet was also explored. In this subsequent paper, we now focus on the above features from winds launched in a magnetized ambient medium. In our studies, the jets and winds are launched in a hot halo at temperatures of $10^{7}~$K equipped with a turbulent magnetic field. Thus, our primary focus is on examining the evolution of jets/winds in an idealized scenario and understanding how this impacts their radio emission characteristics over time. This paper is organized as follows. The simulation setup for the study is briefly discussed in Sec.~\ref{sec:setup}. Detailed results are presented in Sec.~\ref{sec:results}, where we specifically examine the dynamics, emission, and polarization characteristics in separate subsections. We discuss the distinctions in the emission and polarization features of jets from Paper I and winds from this study in Sec.~\ref{sec:jet_wind_compare}. In Sec.~\ref{sec:discussion}, we discuss our findings and summarize our inferences regarding the jets and winds derived from our simulations.

\section{Simulation Setup}
\label{sec:setup}

The setup for the ambient medium and the wind with mildly relativistic speeds in this study is kept similar to Paper I for the jets. Thus, initially, a hot halo is set up which is in the hydrostatic equilibrium in the presence of an external gravitational potential ($\phi(r)$) as,
\begin{equation}
\frac{d p_a}{d r} = - \rho_a (r)\frac{d \phi (r)}{dr}; \qquad p_a(r) = \frac{\rho_a (r)}{\mu m_a} k_B T_h     
\end{equation}
from which we derive
\begin{equation}\label{eq:halo}
p_a(r) = (n_0 k_B T_h) \exp{\left[- \left( \frac{\mu m_a}{k_B T_h}\right) \phi (r) \right]}, T_h = 10^7\,{\rm K}   
\end{equation}
 where $p_a$ and $\rho_a = \mu m_a n_h$ are respectively the pressure and density of the halo gas at spherical radius $r$, $\mu=0.6$ is the mean molecular weight of the fully ionized gas, $m_a$ is the atomic weight, and $n_0$ is the number density at $r=0$. The Boltzmann constant is shown using $k_B$, and $T_h$ is the temperature of the halo. The total gravitational potential ($\phi(r)$) is the sum of the baryonic and dark-matter potential, as described in Paper I in detail. Then, the turbulent magnetic fields are introduced into the ambient medium, and the winds are launched from the lower $Z$ side of the simulation domain.

However, there are several differences, which are explained below. The simulation domain is a large box with dimensions of 8~kpc $\times$ 8~kpc $\times$ 4~kpc, corresponding to a grid of $X \times Y \times Z$, with a resolution of $800 \times 800 \times 400$ cells. As the winds are expected to be broad, the wide box encompasses the cocoon and SAM for the winds by the time its head reaches the $Z$-altitude of 4~kpc. Similar to Paper I, we employ the RMHD module in \textsc{pluto} \citep{mignone_2007} along with a five-wave HLLD Riemann solver \citep{mignone_2009}. We use a piece-wise parabolic scheme for reconstruction \citep{mignone_2005} and a third-order Runge-Kutta method for time integration. In order to ensure numerical instability we use a more diffusive (HLL) Riemann solver and MIN-MOD limiter for shocked cells which are identified using a pressure condition with $\delta p/p_{min} > 6$ in regions near the injection zone ($|X, Y|\leq 0.5$ and $Z\leq 0.5$~kpc).

The turbulent magnetic field in the halo is incorporated by interpolating from a domain of size 4~kpc$^3$, as detailed in Paper I for compact jets. Given the larger simulation box here, the field is extended along the $X$ and $Y$ axes using periodic boundary conditions, as also done in Paper I for large-scale jet run. In this study, the maximum correlation length of the external magnetic field is 1000~pc. The investigation for varying the correlation lengths has already been performed in Paper I, and we anticipate similar outcomes in this regard. Consequently, our primary focus here is on examining the dynamics of winds with different powers launched in a similar turbulent magnetic field configuration. We use the constrained transport approach \citep{balsara_1999, mignone_2021} to maintain the divergence-free nature of the magnetic field throughout the simulation. Additional specifics regarding the setup and numerical schemes can be found in Paper I, to which we direct the reader for further details.

The simulations performed in this study are listed in Table~\ref{tab:sim_table}. The nomenclature of the simulations indicates the power of the injected wind and the correlation length of the ambient magnetic field, as explained in the footnote of Table~\ref{tab:sim_table}. The half-cone opening angle for the wind ($\theta_W$) is set at $70^\circ$ for a broader injection than the jets, which were launched with an angle of $7^\circ$. The radius of the injection zone is 0.2~kpc here; thus the diameter for the injected wind is resolved with 40 cells for a resolution of 10~pc in our runs. Simulations with high-density contrast ($\eta_w$) between the wind and the ambient medium ($\mathrm{W43\_L1000(D,HD)}$) are also performed to explore the effect of a dense wind on different physical measures. We also conduct runs for narrow winds, which are launched with a half-cone opening angle ($\theta_w$) of $35^\circ$, and explore this for both light and dense wind cases ($\mathrm{W43\_L1000(N,HDN)}$). It can be seen that the winds exhibit mildly relativistic velocities in comparison to the jets in Paper I, which were launched with speeds reaching up to 0.99c. Outflows with velocities of around 0.25c, driven off by the accretion disc winds at distances of a few gravitational radii from the black hole, have been observed in the nuclear regions of several quasars \citep[see e.g.][]{pounds_2003, braito_2007, reeves_2009,tombesi_2015}. These winds become mildly relativistic with distance and are likely to attain velocities similar to the values chosen in our study. For post-process analysis, we consider an electron population whose energy spectrum follows a power law with an index of 2.2, similar to the one assumed in Paper I. The synchrotron emissivities from radiation emitted by the acceleration of electrons in the presence of a magnetic field are estimated using the expressions given in Sec.~3 of Paper I. In the next section, we discuss the results of the dynamics and observable characteristics of winds from this study.

\begin{table*}
	\centering
	\caption{Parameters for simulations performed in this study.}
	\label{tab:sim_table}
	\begin{threeparttable}
	
	\begin{tabular}{|c|l|l|l|l|l|l|l|} 
		\hline
Simulation & $\eta_w$ & $V_w$ & $P_w [\mathrm{\ergs}]$ &  $B_0 [\mathrm{mG}]$ & $M_w$ \\
		\hline
		$\mathrm{W42\_L1000}$ & $4 \times 10^{-2}$ &  0.021c & $1.01 \times 10^{42}$ & 0.033 & 2.68 \\
		\hline
			$\mathrm{W43\_L1000}$ & $4 \times 10^{-2}$ &  0.051c & $1.03 \times 10^{43}$ & 0.069 & 6.36 \\
	\hline
	$\mathrm{W44\_L1000}$ & $4 \times 10^{-2}$ & 0.11c & $1.04 \times 10^{44}$ & 0.148 & 14.02\\	
		\hline
  $\mathrm{W43\_L1000D^a}$ & $4 \times 10^{-1}$ &  0.024c & $1.04 \times 10^{43}$ & 0.10 & 9.5 \\ 
  \hline
  $\mathrm{W43\_L1000HD^a}$ & $4$ &  0.011c & $1.06 \times 10^{43}$ & 0.148 & 14.11 \\ 
  \hline
	 $\mathrm{W43\_L1000N^b}$ & $4 \times 10^{-2}$ &  0.056c & $1.02 \times 10^{43}$ & 0.076 & 7.1 \\
  \hline
   $\mathrm{W43\_L1000HDN^b}$ & 4 &  0.011c & $0.8 \times 10^{43}$ & 0.15 & 14.11 \\
		\hline
	\end{tabular}
	\begin{tablenotes}
 \item The simulation label indicates both the wind power and the correlation length of the ambient magnetic field. For example, $\mathrm{W42\_L1000}$ represents a simulation where a wind with a power of approximately $10^{42}~\ergs$ is launched into an ambient medium with a mean magnetic field of about 10 $\mu$G and a correlation length of 1000 pc.
  \item[a] Dense (D, $\eta_w = 4 \times 10^{-1}$), and highly dense (HD, $\eta_w = 4$) wide-angled winds of power $10^{43}~\ergs$.
 \item[b] Light and narrow (N, $\theta_w=35^\circ$) and Highly dense and narrow (HDN, $\eta_w = 4$ and $\theta_w=35^\circ$) winds of power $10^{43}~\ergs$.\\
$\eta_w$: Ratio of wind density to the ambient gas density at the radius of injection. \\
 $V_w$: Bulk velocity of the wind.\\
 $P_w$: Mechanical power of the wind.\\
 $B_0 $: Maximum strength of the toroidal magnetic field of the wind at the injection zone. Note that the magnetization of the wind ($\sigma$) as defined in Paper I (Eq.~7) is taken as 0.1 for all the simulations.\\
$M_w$: Mach number of the wind.\\

	\end{tablenotes}
	\end{threeparttable}
\end{table*}

\begin{figure*}
\centerline{ 
\def\arraystretch{1.0}
\setlength{\tabcolsep}{0.0pt}
\begin{tabular}{lcr}
  \includegraphics[angle=0,origin=c,width=14cm,keepaspectratio]{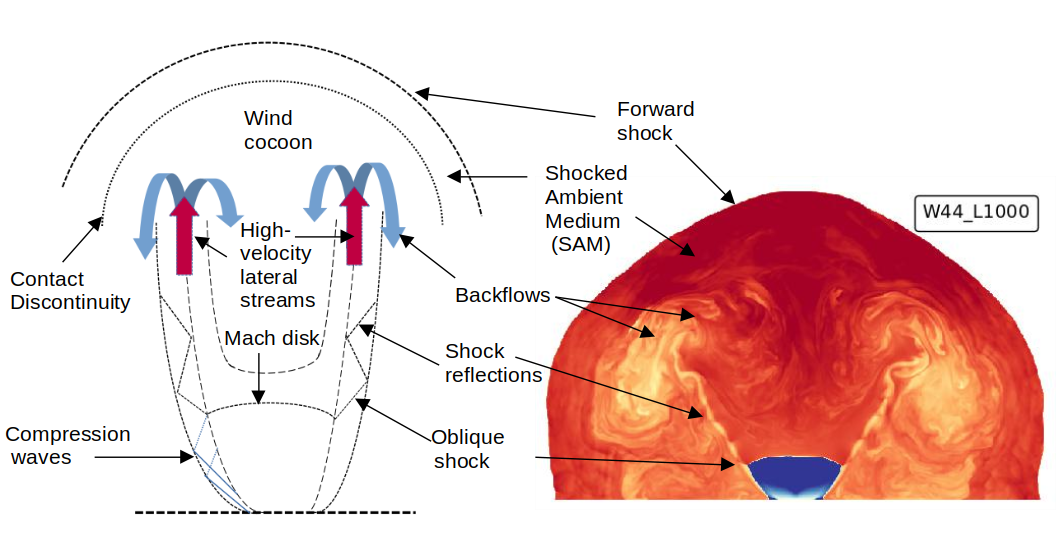}
    \end{tabular}}
       \caption{Schematic diagram showing the structure of wind (left) and pressure map for $\mathrm{W44\_L1000}$ (right) for illustration (similar to Fig.~\ref{fig:rho_prs_tr1}).}
\label{fig:wind_structure}
\end{figure*}

\begin{figure*}
\centerline{ 
\def\arraystretch{1.0}
\setlength{\tabcolsep}{0.0pt}
\begin{tabular}{lcr}
  
  \includegraphics[scale=0.5,keepaspectratio]{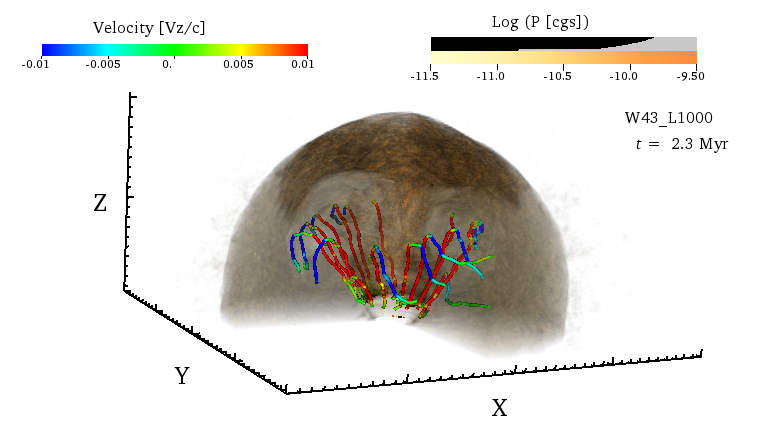}
    \end{tabular}}

       \caption{3D volume rendering of logarithmic pressure and velocity streamlines showing the trajectories of fluid in the cocoon for $\mathrm{W43\_L1000}$.}
\label{fig:3d}
\end{figure*}

\begin{figure*}
\centerline{ 
\def\arraystretch{1.0}
\setlength{\tabcolsep}{0.0pt}
\begin{tabular}{lcr}
  \includegraphics[scale=0.33,keepaspectratio]{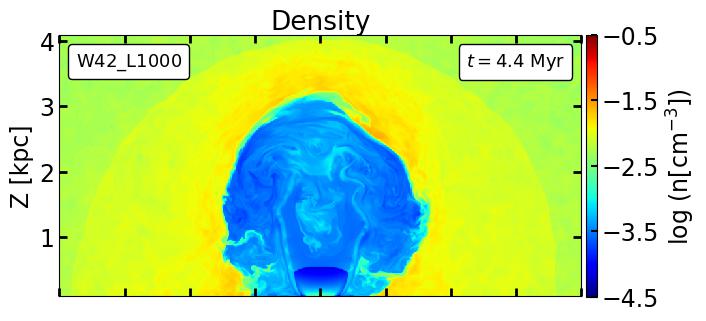}
    \includegraphics[scale=0.33,keepaspectratio]{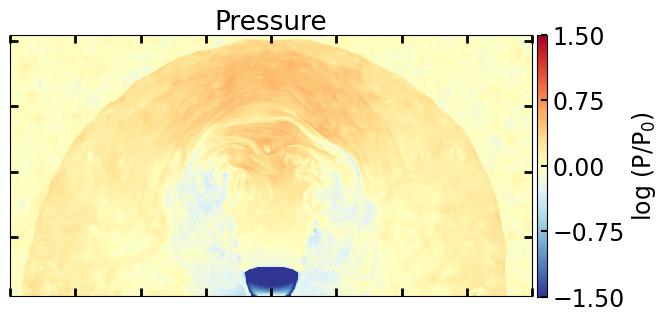}
    \includegraphics[scale=0.33,keepaspectratio]{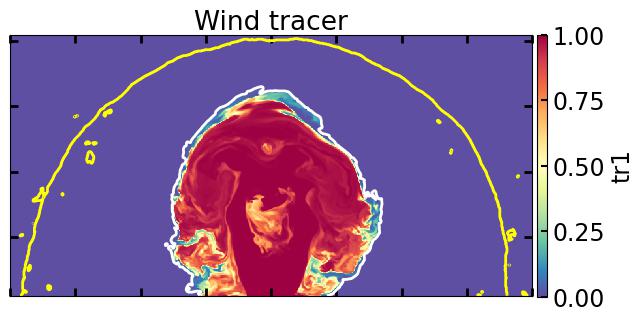}
    \end{tabular}}

    \centerline{ 
\def\arraystretch{1.0}
\setlength{\tabcolsep}{0.0pt}
\begin{tabular}{lcr}
  \includegraphics[scale=0.33,keepaspectratio]{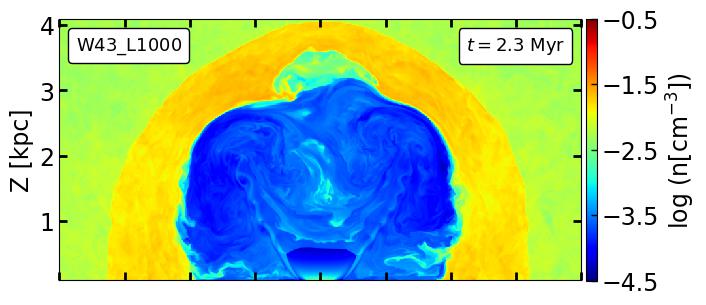}
    \includegraphics[scale=0.33,keepaspectratio]{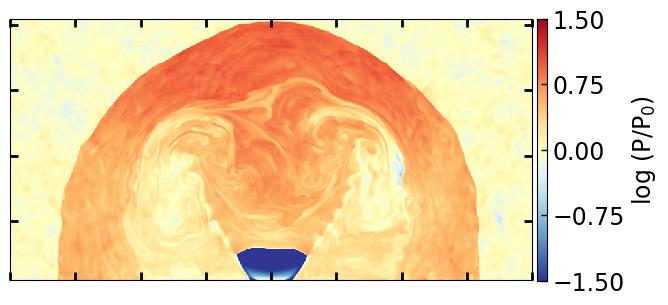}
    \includegraphics[scale=0.33,keepaspectratio]{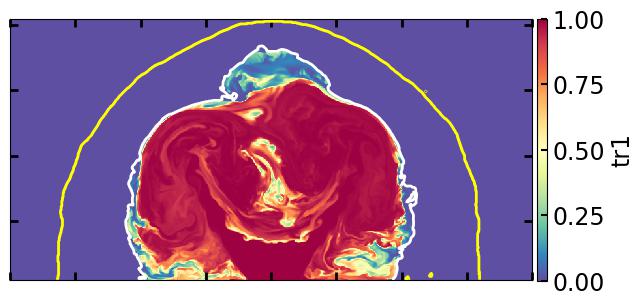}
    \end{tabular}}

    \centerline{ 
\def\arraystretch{1.0}
\setlength{\tabcolsep}{0.0pt}
\begin{tabular}{lcr}\hspace{0.01cm}
  \includegraphics[scale=0.33,keepaspectratio]{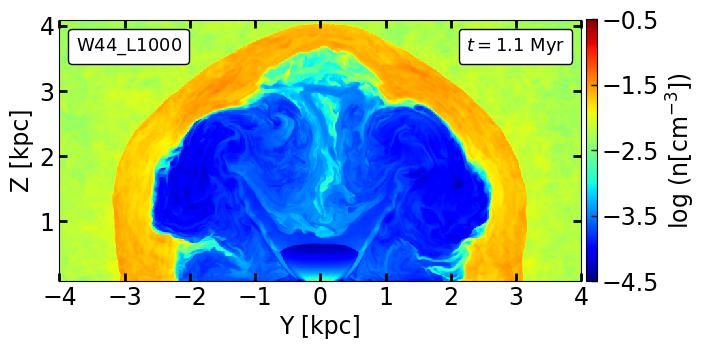}\hspace{-0.15cm}
    \includegraphics[scale=0.33,keepaspectratio]{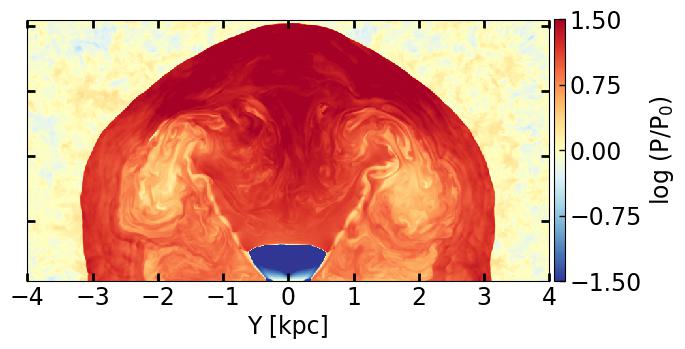}\hspace{-0.15cm}
    \includegraphics[scale=0.33,keepaspectratio]{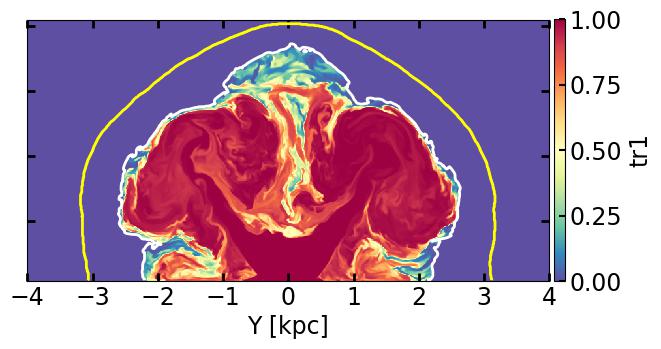}
    \end{tabular}}
       \caption{Logarithmic density (left), pressure (middle), and wind tracer (right) for simulations of wide-angled ($\theta_W=70^\circ$) light winds ($\eta_w=4\times 10^{-2}$) of different powers ($10^{42}-10^{44}~\ergs$) in the $Y-Z$ plane. The region enclosed inside the cocoon and forward shock (see Appendix~\ref{forward_shock}) are represented by white and yellow contours in the right panels.}
\label{fig:rho_prs_tr1}
\end{figure*}

\begin{figure*}
\centerline{ 
\def\arraystretch{1.0}
\setlength{\tabcolsep}{0.0pt}
\begin{tabular}{lcr}
  \includegraphics[scale=0.335,keepaspectratio]{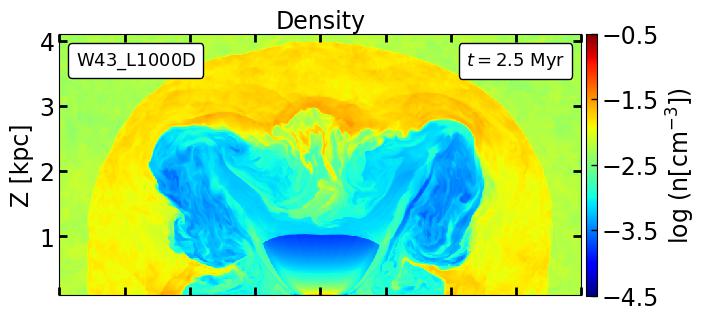}
    \includegraphics[scale=0.335,keepaspectratio]{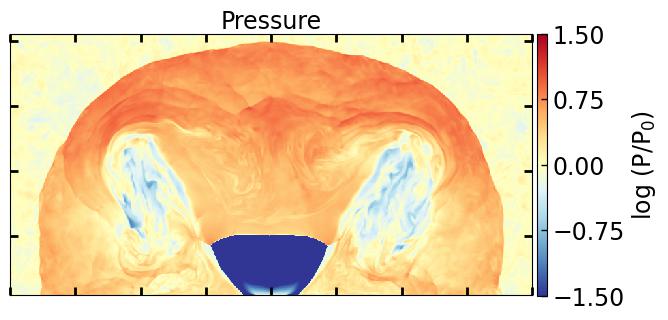}
    \includegraphics[scale=0.335,keepaspectratio]{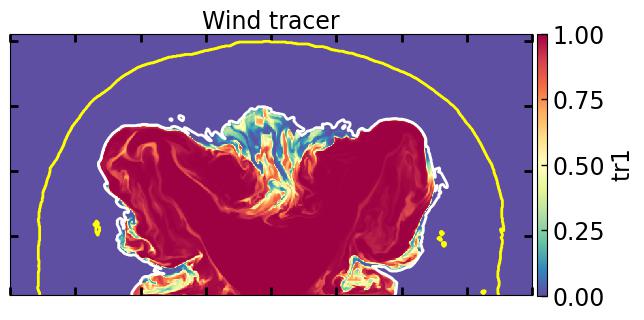}
    \end{tabular}}

\centerline{ 
\def\arraystretch{1.0}
\setlength{\tabcolsep}{0.0pt}
\begin{tabular}{lcr}
  \includegraphics[scale=0.33,keepaspectratio]{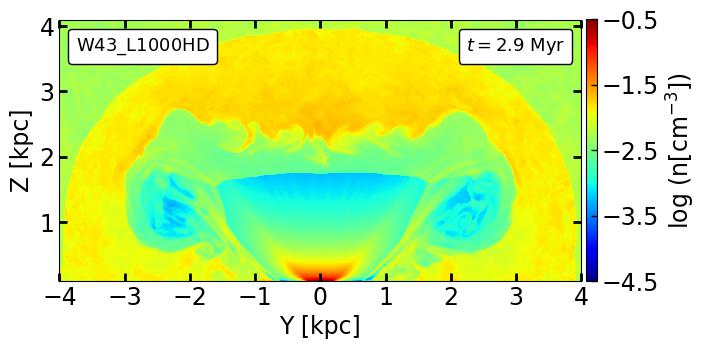}
    \includegraphics[scale=0.33,keepaspectratio]{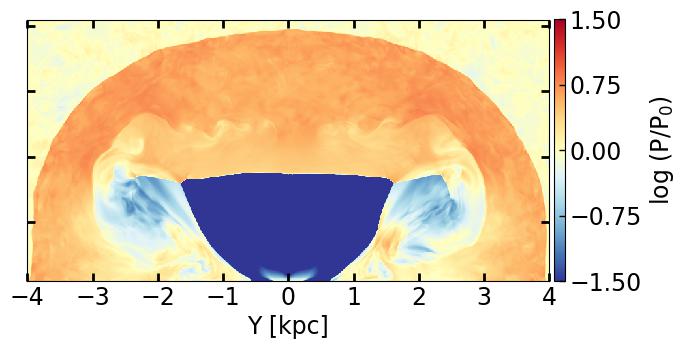}
    \includegraphics[scale=0.33,keepaspectratio]{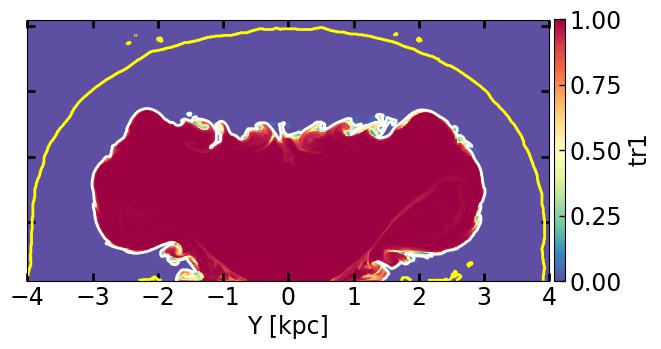}
    \end{tabular}}

       \caption{Logarithmic density (left), pressure (middle), and wind tracer (right) for the wide-angled dense ($\eta_w=4\times 10^{-1}$; top row) and highly dense wind ($\eta_w=4$; bottom row) with power $10^{43}~\ergs$. The region enclosed inside the cocoon and forward shock (see Appendix~\ref{forward_shock}) are represented by white and yellow contours in the right panels.}
\label{fig:rho_43_n2}
\end{figure*}


\begin{figure*}
\centerline{ 
\def\arraystretch{1.0}
\setlength{\tabcolsep}{0.0pt}
\begin{tabular}{lcr}
  \includegraphics[scale=0.335,keepaspectratio]{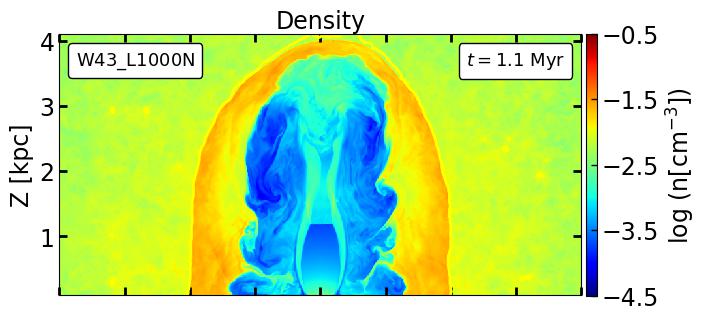}
    \includegraphics[scale=0.335,keepaspectratio]{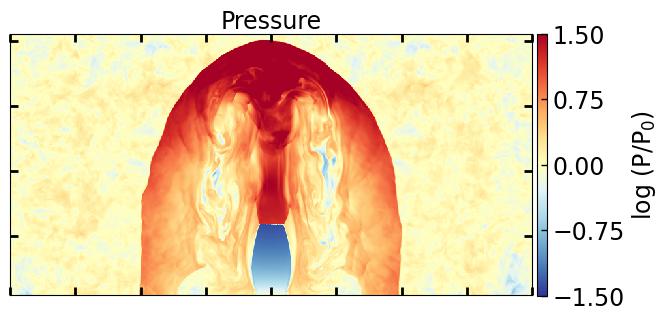}
    \includegraphics[scale=0.335,keepaspectratio]{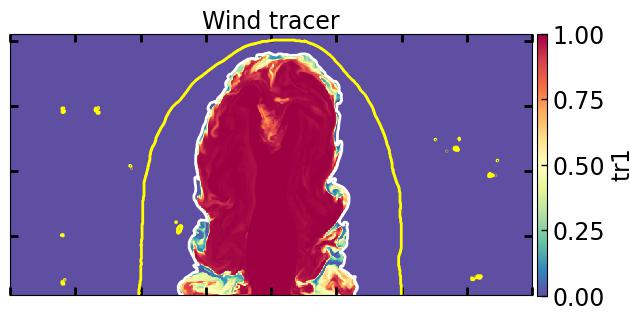}
    \end{tabular}}
       
\centerline{ 
\def\arraystretch{1.0}
\setlength{\tabcolsep}{0.0pt}
\begin{tabular}{lcr}

    \includegraphics[scale=0.33,keepaspectratio]{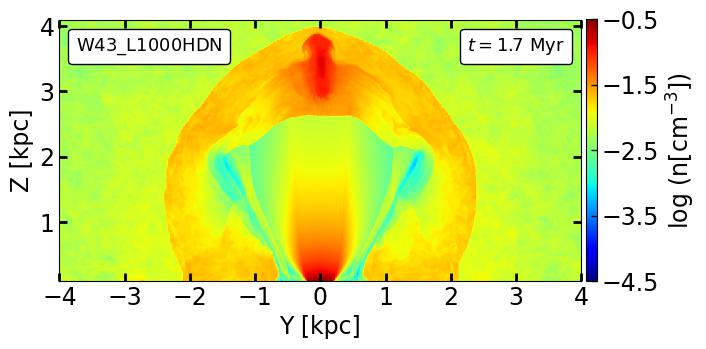}  
    \includegraphics[scale=0.33,keepaspectratio]{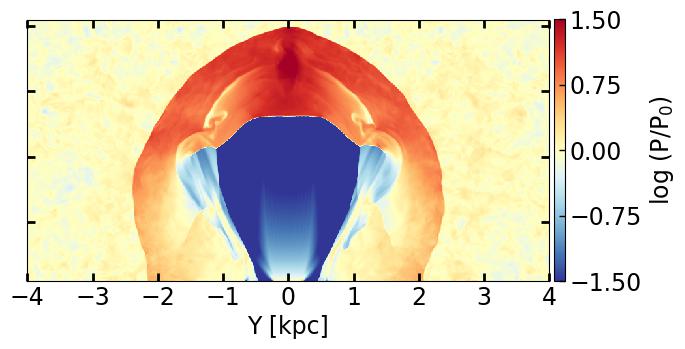}
    \includegraphics[scale=0.33,keepaspectratio]{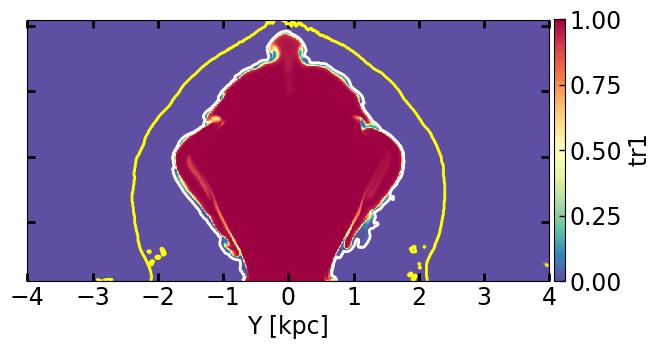}
    \end{tabular}}

      \caption{Logarithmic density (left), pressure (middle), and wind tracer (right) for the light ($\eta_w=4\times 10^{-2}$; top row) and highly dense ($\eta=4$; bottom row) winds with a narrow opening angle ($\theta_w=35^\circ$). The winds have a power of $10^{43}~\ergs$. The region enclosed inside the cocoon and forward shock (see Appendix~\ref{forward_shock}) are represented by white and yellow contours in the right panels.}
\label{fig:rho_43_n2_th}
\end{figure*}

\begin{figure*}
\centerline{ 
\def\arraystretch{1.0}
\setlength{\tabcolsep}{0.0pt}
\begin{tabular}{lcr}
  \includegraphics[scale=0.39,keepaspectratio]{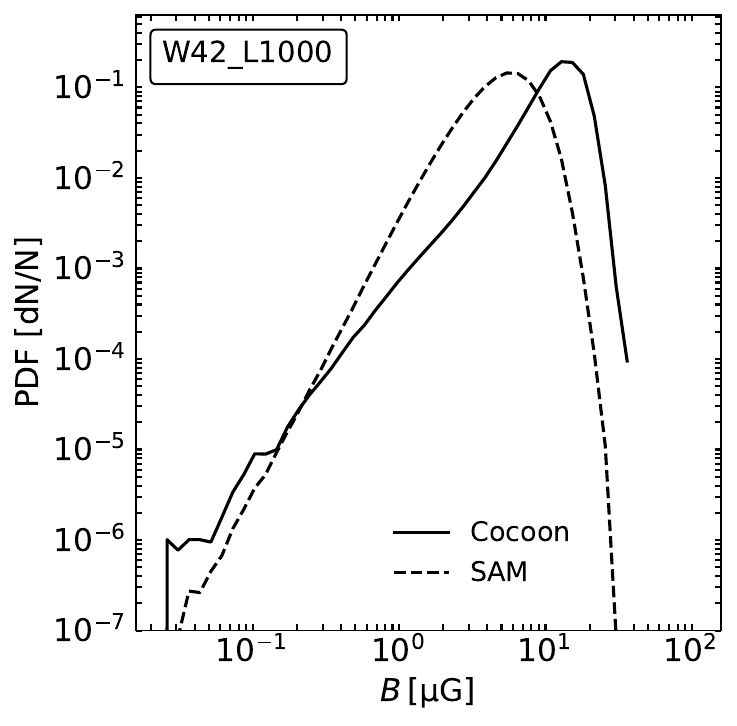}
    \includegraphics[scale=0.416,keepaspectratio]{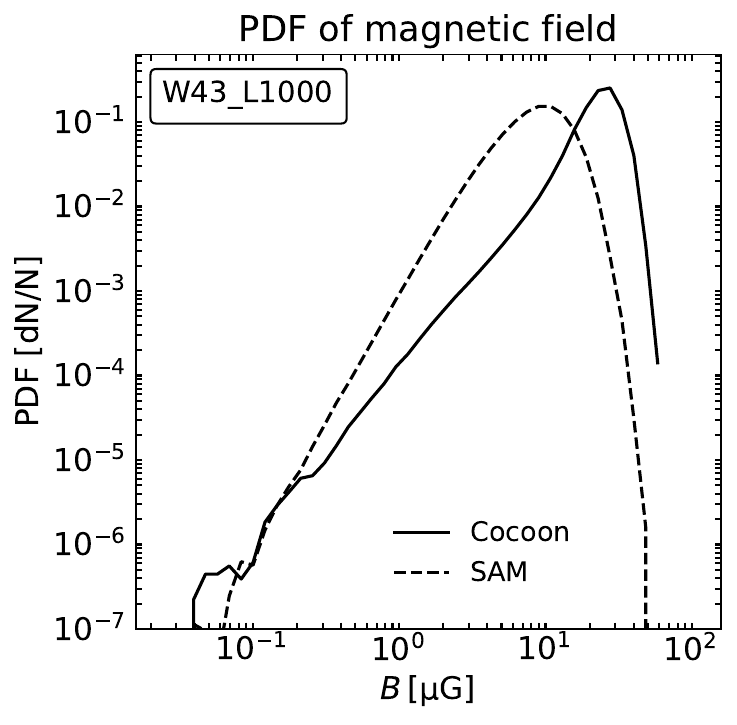}
    \includegraphics[scale=0.39,keepaspectratio]{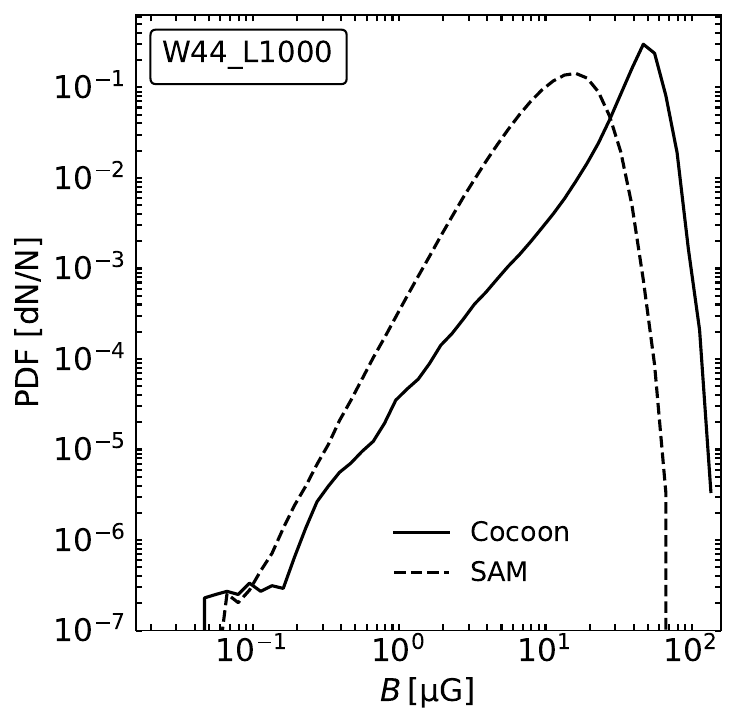}
    \end{tabular}}
       \caption{Probability Distribution of magnetic field in the cocoon ($\mathrm{tr1}>10^{-3}$) and SAM (defined in Appendix~\ref{forward_shock}) for light winds W42, W43 and W44. The times of the snapshots are similar to those in Fig.~\ref{fig:rho_prs_tr1}, corresponding to when the head of the forward shock reaches close to the top of the simulation domain.}
\label{fig:dist_B}
\end{figure*}

\section{Results} 
\label{sec:results}
\subsection{Dynamics of wind}
\label{sec:dynamics}
In this section, we discuss the structure of the wind, and how its characteristics vary in response to various factors, including wind power ($P_w$), density contrast between the injected wind and the ambient medium ($\eta_w$), and the wind's opening angle ($\theta_w$). The general structure of light wind is illustrated through a schematic diagram and a 3D volume-rendered pressure map in Fig.~\ref{fig:wind_structure} and~\ref{fig:3d}, respectively. Slices depicting density, pressure, and wind tracer\footnote{Tracers are passive scalar quantities between 0 and 1 in \textsc{pluto}, which gets advected with the fluid. The wind tracer (tr1) value at the injection zone is set to 1 at $t=0$, and zero elsewhere. See \url{http://plutocode.ph.unito.it/userguide.pdf} for further clarification.} in the $Y-Z$ plane from different simulations are shown in Figs.~\ref{fig:rho_prs_tr1},~\ref{fig:rho_43_n2}, and~\ref{fig:rho_43_n2_th}. These snapshots correspond to the times when the front head of the forward shock reaches close to the top of the simulation domain; indicating that the spatial scale of the winds is similar across all simulations.

\begin{itemize}
    \item \textbf{Structure of the wind:} The general structure of the wind is expected to align with the model presented in \citet{weaver_1977} and \citet{faucher_2012}. However, our study reveals deviations in the wind morphology, which primarily depend on the initial density ratio between the wind and the ambient medium ($\eta_w$). Specifically, dense winds ($\eta_w \gtrsim 4$ in this study) are stable and closely adhere to the wide-angle wind model, as depicted in above studies. Conversely, light winds ($\eta_w \lesssim 0.4$) are more susceptible to external influences, resulting in instabilities and some deviations from this structure. Therefore, as the density of the wind decreases over time due to expansion, instabilities are likely to emerge. In this section, we focus primarily on the structure of the light wind, detailing the instabilities. A labeled schematic diagram for the structure of the light wind is shown in Fig.~\ref{fig:wind_structure}. Adjacent to it, the pressure map for $\mathrm{W44\_L1000}$ is presented to illustrate the various elements. The main features inside the wind cocoon include a Mach disk illustrating the wind's termination shock, and lateral streams emerging from the edges of the barrel (or conical-shaped region), which are shown using red arrows. The wind plasma is abruptly stopped at the termination shock of the Mach disk. However, the oblique shock near the outer edges of the barrel deflects the gas moving downstream. This leads to high-velocity lateral streams emerging around the barrel, which can also be seen from the 3D volume rendering of the pressure along with fluid velocity streamlines in Fig.~\ref{fig:3d}. The shock reflections can be seen in the two streams in the pressure map in Fig.~\ref{fig:wind_structure}. Similar features can also be seen for other cases in Fig.~\ref{fig:rho_prs_tr1}. This wind structure bears similarity to the under-expanded jets \citep[e.g.][]{jennifer_2012} and has also been observed in the previous simulations of wide-angled winds/jets \citep[e.g. see][]{krause_2012,wagner_2013,santanu_2022}.

  The backflows from the lateral streams are directed inwards i.e. in regions between the streams, as well as in the outward direction, as depicted using blue arrows in Fig.~\ref{fig:wind_structure}. Generally, the inward-directed backflows are weaker compared to the outward ones. As can be seen from the velocity streamlines shown in Fig.~\ref{fig:3d}, these backflows lead to eddies/vortices in the cocoon. The eddies lead to velocity shears in the inner layers of the contact discontinuity, thereby facilitating some entrainment (or mixing) of the shocked ambient gas with the wind plasma. This causes wind tracer (tr1) values lower than 1 in these regions, which will be discussed in the next sections.
    
        \item \textbf{Dependence on the wind power:} The logarithmic density, pressure, and wind tracer maps for simulations with different wind powers are shown in Fig.~\ref{fig:rho_prs_tr1}. The white and yellow contours in the right panels show the regions enclosed within the cocoon and forward shock, respectively. It can be seen that as the wind power decreases, the forward shock takes a more spherical shape. Conversely, high-power winds carry strong forward momentum, resulting in an elongated appearance along the $Z$ direction. The SAM for W42 shows a double-layered structure where the density is very high in regions closer to the cocoon, and decreases outwards. The inner SAM consists of the old swept-up ambient medium and hence depicts high gas density. The outer SAM also indicates that the forward shock from W42 is not as powerful as the high-power winds by the time it reaches a similar altitude. This can affect the acceleration of electrons at the forward shock and in turn, the polarized emission, which is discussed later in Sec.~\ref{sec:synch_pol_vary}.

   The backflows on the lateral sides of the cocoon generate eddies (as shown in Fig.~\ref{fig:3d}), leading to a reduction in gas pressure in these regions. Simultaneously, a high-pressure region develops in the forward zone ahead of the Mach disk. The outward eddies act to confine the wind, resulting in collimation and instabilities in the fast streams from the edges of the barrel. Inward-directed backflows and high-pressure zones ahead of the Mach disk, as observed in W43 and W44, counteract this collimation. However, this is not the case for W42 due to weak inward-directed backflows. This suggests that low-power, light winds are likely to experience strong collimation from the surroundings, resulting in a considerably narrower cocoon, and conversely for the high-power ones. 

The wind tracer maps in the right panels indicate that the cocoons undergo some mixing with the ambient gas. The tracer values are 1 inside the conical zone and along the fast lateral streams. High tracer values (tr1$>0.75$) are also observed in the central and lateral regions of the cocoon, where the backflowing wind plasma is present. Some mixing can be seen on the lateral sides, and is significant near the cocoon head, resulting in lower wind tracer values (tr1$\lesssim0.25$). Initially, the formation of fast streams and strong backflows in the lateral parts of the cocoon causes the ambient gas to fill most regions between the streams above the Mach disk. Inward-directed backflows cause shearing in the inner layers of SAM, leading to some entrainment of the external medium. This results in low wind tracer values near the top head of the cocoon. Stronger backflows in high-power winds cause more entrainment compared to low-power winds. The low-power wind W42 initially confines itself within a narrow cocoon, limiting the extent of its eddies and reducing mixing with the external medium.
    
    \item \textbf{Dependence on the density of the wind:} Maps for different physical quantities for a dense and highly dense wind with a power of $10^{43}\ergs$ are presented in Fig.~\ref{fig:rho_43_n2} (referred to as $\mathrm{W43\_L1000D}$ and $\mathrm{W43\_L1000HD}$ in Table~\ref{tab:sim_table}, respectively). It is evident that, as the wind density increases, the Mach disk in the winds is able to advance more along the $ Z-$direction. This is because the dense winds carry a high momentum flux and are, thus, weakly affected by interaction with the surroundings. Consequently, the lateral streams manifest at later times when the wind has progressed further than the case of the light wind in Fig.~\ref{fig:rho_prs_tr1}, resulting in shorter streams. This implies that with increasing density, the winds become more stable, closely resembling the wide-angled wind model from \citet{faucher_2012}.
    
   It can also be seen that the cocoon of dense winds lacks the characteristic ``hump-like'' feature at the head, as observed in the case of light winds in Fig.~\ref{fig:rho_prs_tr1}. While a small ``hump'' is discernible in the dense wind, it is notably absent in the highly-dense case. In light winds, the lateral streams are unstable due to confinement force from eddies in the lateral regions, and thus the backflowing plasma is directed both inwards and outwards.  The inward-directed backflows try to fill the volume ahead of the Mach disk, resulting in some plasma being pushed upward, creating a distinct ``hump-like'' structure within the cocoon. In contrast, when the wind is dense, and thus, relatively stable, the lateral streams emanating from the edges of the barrel are comparatively stable. As a result, strong backflows from high-velocity regions occur primarily along the lateral sides and are weaker inwards, leading to the absence of a prominent hump at the cocoon's head.

   The tracer maps displayed in the right panel indicate that some entrainment of the external gas (tr1$\lesssim0.25$) is present near the cocoon head of the dense wind. This results from the formation of fast lateral streams and inward-directed backflows at later stages. However, in the case of the highly dense wind, the lateral streams are still in their initial phase, and thus no entrainment of the shocked ambient gas is seen here.
    
     \item \textbf{Dependence on the opening angle of the wind:} Maps for a light and highly dense wind with a half-cone opening angle ($\theta_W$) of $35^\circ$ and power $10^{43}~\ergs$ are shown in Fig.~\ref{fig:rho_43_n2_th}. It can be seen that a narrow light wind experiences strong collimation from its surroundings when compared to the wide-angled wind of similar power in Fig.~\ref{fig:rho_prs_tr1}. Consequently, the wind becomes more constrained, resulting in the formation of a slender, elongated high-pressure zone ahead of the Mach disk. This also leads to high-pressure knots in these regions and a high-pressure cocoon head. Similar high pressure can also be seen near the cocoon head of the dense narrow wind, resulting from collimation from the surroundings. The dense wind is stable than the light wind here, and so fast lateral streams have not yet developed. The narrow winds undergo weak mixing with the external medium, as can be seen from the tracer map in the right panel.
     
\end{itemize}

Similar to the jets in Paper I, the bow shock of the wind compresses the external magnetic fields. Fig.~\ref{fig:dist_B} displays the probability density function (PDF) of the magnetic field within the cocoon and SAM across wind simulations with different powers. In all cases, the PDF for the SAM consistently lies on the left side of the cocoon, indicating comparatively weaker magnetic fields. However, it is noteworthy that the peaks in the PDF from the cocoon and SAM get close in magnitude as the wind power is decreased. This discrepancy arises due to the decay of the magnetic field within both the cocoon and the SAM as expansion occurs, as previously shown in Paper I. Consequently, the magnetic field within the cocoon of low-power winds, which are initiated with comparatively weaker magnetic fields (refer to Table~\ref{tab:sim_table}), has significantly decayed to values comparable to those within the SAM. Conversely, the magnetic fields in the high-power winds, particularly for W44, are sustained at values higher than the SAM by the time the cocoon's head reaches the top of the simulation domain. The distinct magnetic field strengths in the cocoon and SAM lead to varying effects on the emission and polarization originating from the wind, as elaborated in the subsequent sections.

\subsection{Synchrotron emission and polarization from winds}
\label{sec:synch_pol}
\begin{figure*}
\centerline{ 
\def\arraystretch{1.0}
\setlength{\tabcolsep}{0.0pt}
\begin{tabular}{lcr}
\includegraphics[scale=0.58,keepaspectratio]{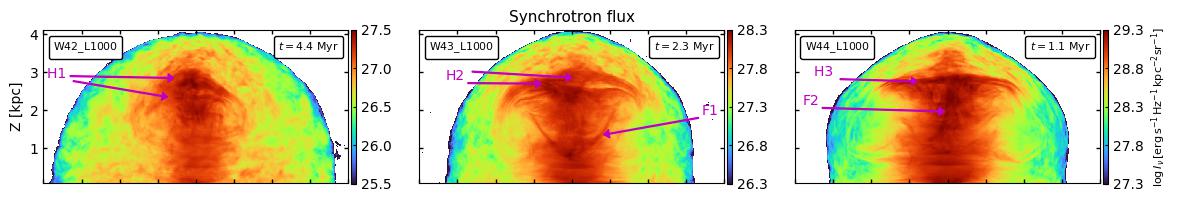}
    \end{tabular}}
    \centerline{ 
\def\arraystretch{1.0}
\setlength{\tabcolsep}{0.0pt}
    \begin{tabular}{lcr}
    \includegraphics[scale=0.58,keepaspectratio]{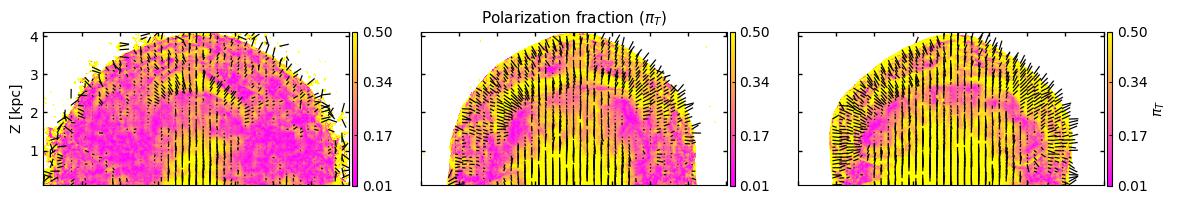}

    \end{tabular}}
        \centerline{ 
\def\arraystretch{1.0}
\setlength{\tabcolsep}{0.0pt}
    \begin{tabular}{lcr}\hspace{0.1cm}
    \includegraphics[scale=0.58,keepaspectratio]{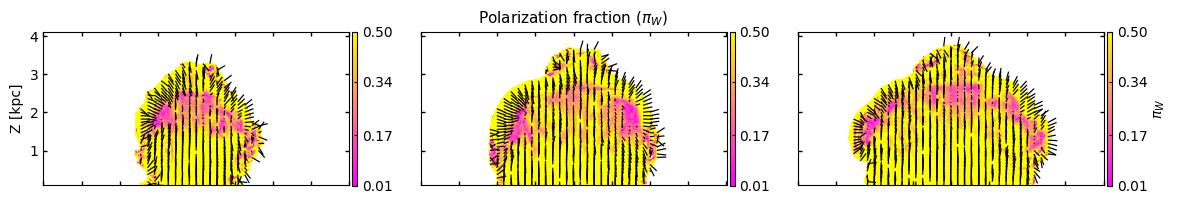}
    \end{tabular}}

      \centerline{ 
\def\arraystretch{1.0}
\setlength{\tabcolsep}{0.0pt}
    \begin{tabular}{lcr}\hspace{0.1cm}
    \includegraphics[scale=0.58,keepaspectratio]{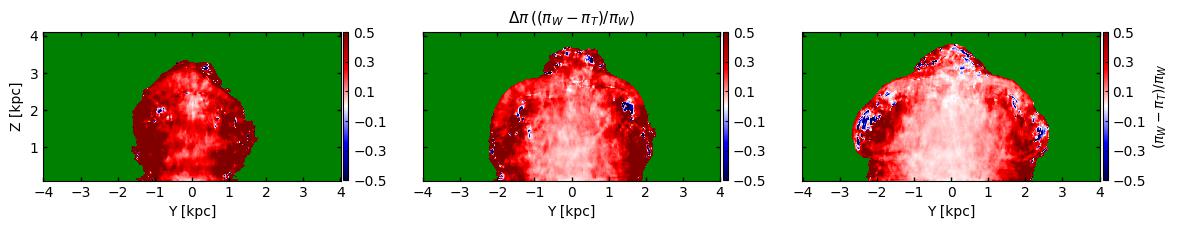}

    \end{tabular}}
       \caption{Logarithmic synchrotron flux (first row), total and cocoon polarization fraction (second and third row), and the fractional change in polarization (Fourth row) at $\theta_I=90^\circ$ image plane for different simulations. H1, H2, and H3 show horizontal bright arcs near the cocoon's head. F1 and F2 show filamentary-like structures resulting from the shearing of magnetic fields.}
\label{fig:diff_power_90deg}
\end{figure*}

\begin{figure*}
\centerline{ 
\def\arraystretch{1.0}
\setlength{\tabcolsep}{0.0pt}
\begin{tabular}{lcr}
  \includegraphics[scale=0.290,keepaspectratio]{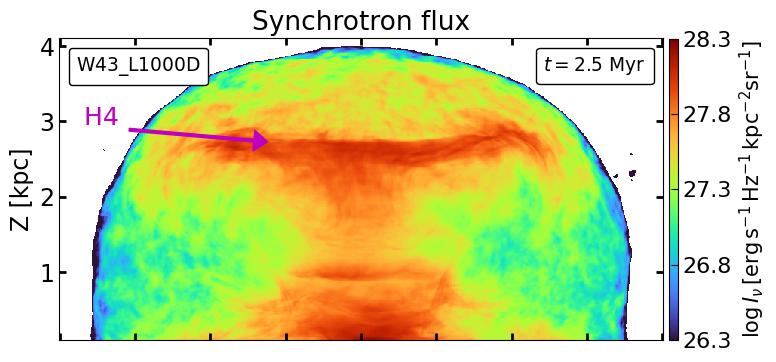}
    \includegraphics[scale=0.280,keepaspectratio]{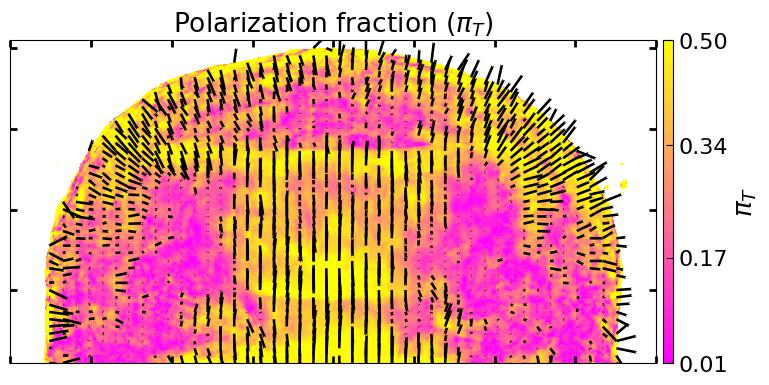}
    \includegraphics[scale=0.280,keepaspectratio]{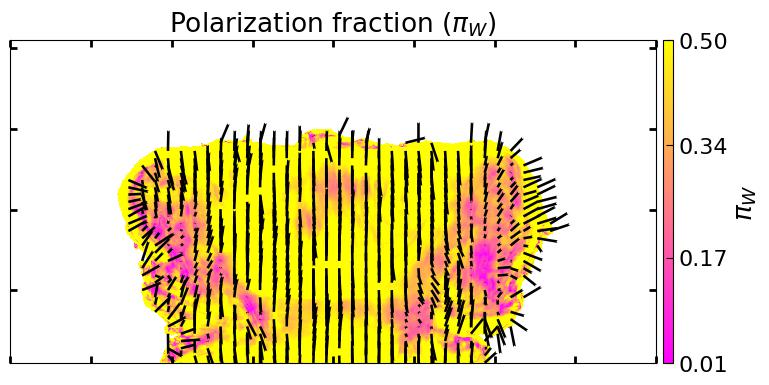}
    \end{tabular}}

\centerline{ 
\def\arraystretch{1.0}
\setlength{\tabcolsep}{0.0pt}
\begin{tabular}{lcr}
  \includegraphics[scale=0.290,keepaspectratio]{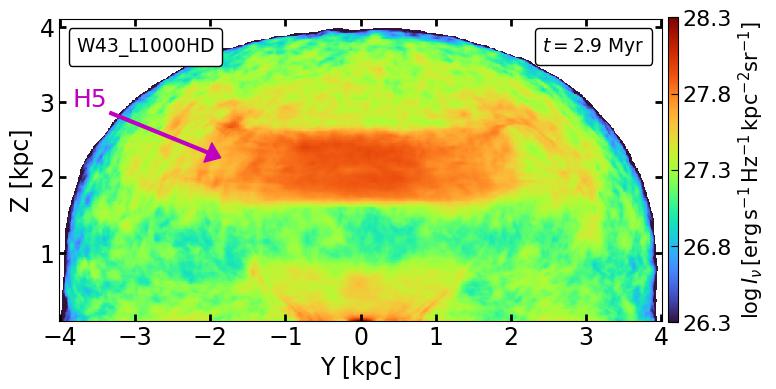}
    \includegraphics[scale=0.280,keepaspectratio]{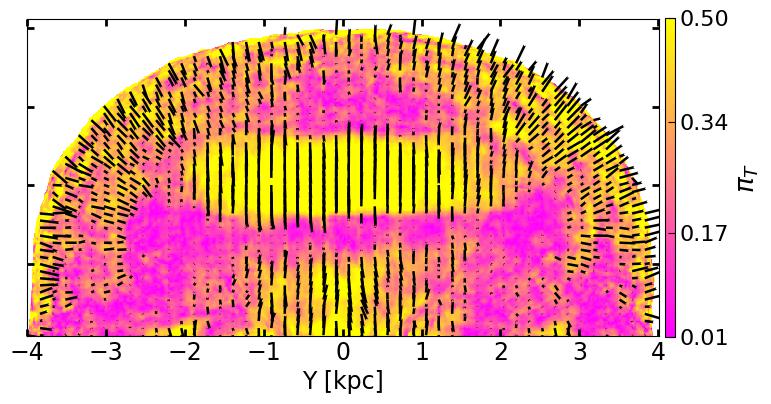}
    \includegraphics[scale=0.280,keepaspectratio]{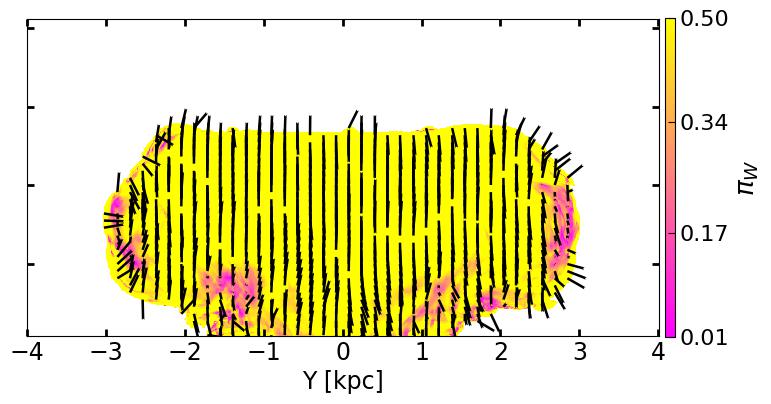}
    \end{tabular}}
 
        \caption{Top: Logarithmic synchrotron flux (left), total and cocoon polarization fraction middle and right) at $\theta_I=90^\circ$ image plane for dense wind case ($\mathrm{W43\_L1000D}$). H4 shows a horizontal bright arc near the cocoon's head. Bottom: Same as above for highly dense wind ($\mathrm{W43\_L1000HD}$). H5 shows a high-emission extended region near the cocoon's head.}
\label{fig:emiss_n2_90deg}
\end{figure*}

\begin{figure*}
    \centerline{ 
\def\arraystretch{1.0}
\setlength{\tabcolsep}{0.0pt}
\begin{tabular}{lcr}
  \includegraphics[scale=0.290,keepaspectratio]{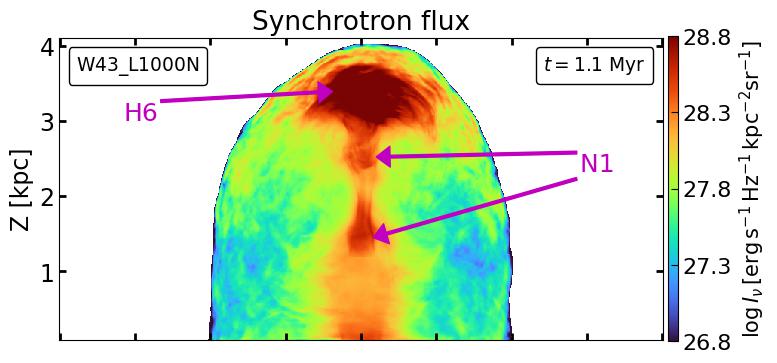}
    \includegraphics[scale=0.280,keepaspectratio]{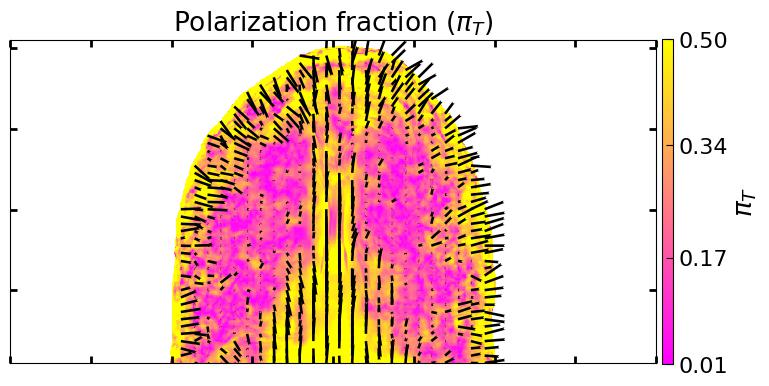}
    \includegraphics[scale=0.280,keepaspectratio]{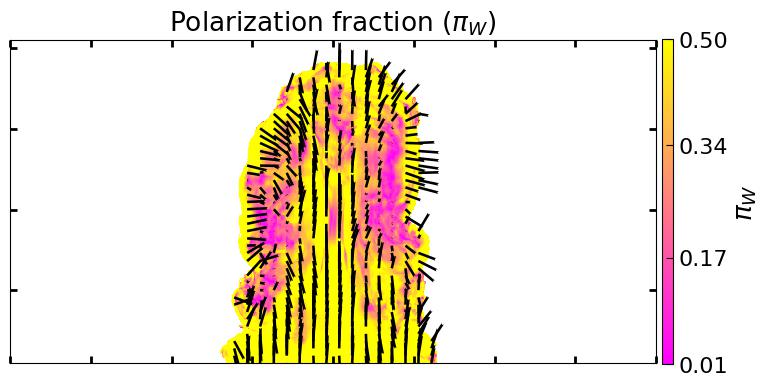}
    \end{tabular}}
\centerline{ 
\def\arraystretch{1.0}
\setlength{\tabcolsep}{0.0pt}
\begin{tabular}{lcr}
\includegraphics[scale=0.290,keepaspectratio]{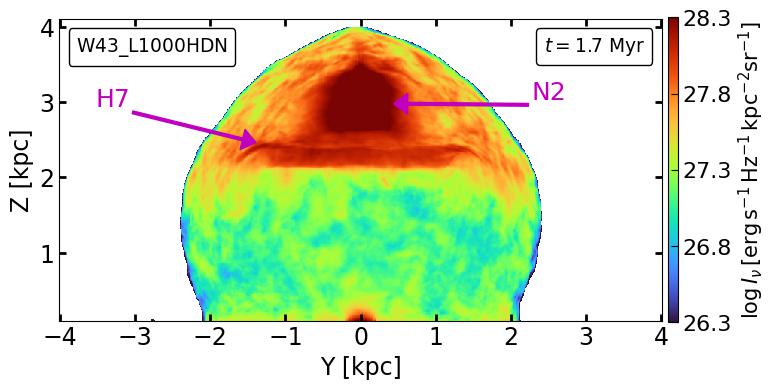}  
    \includegraphics[scale=0.280,keepaspectratio]{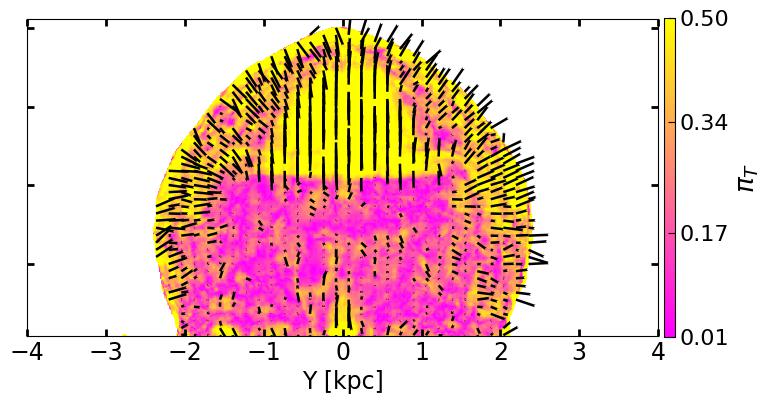}
    \includegraphics[scale=0.280,keepaspectratio]{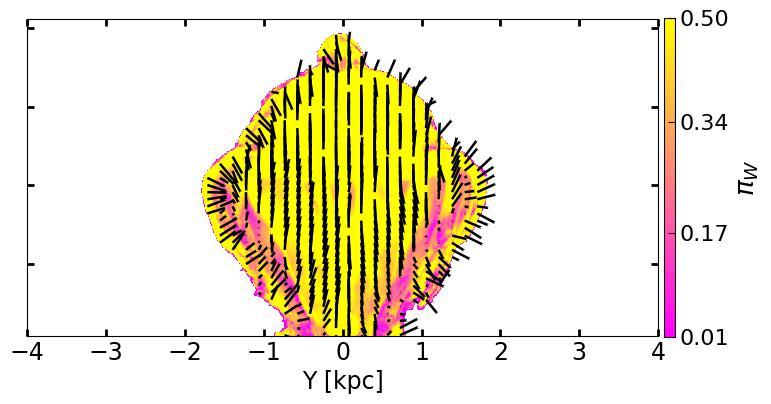}
    \end{tabular}}

       \caption{Top: Logarithmic synchrotron flux (left), total and cocoon polarization fraction (middle and right) at $\theta_I=90^\circ$ image plane for narrow light wind case ($\mathrm{W43\_L1000N}$). H6 shows a high-emission region near the cocoon's head, and N1 are the high-pressure knots that form due to the strong collimation of the wind. Bottom: Same as above for highly dense narrow wind ($\mathrm{W43\_L1000HDN}$). A bright compact region (N2) can be seen above the horizontal bright arc (H7).}
\label{fig:emiss_n2_th_90deg}
\end{figure*}

\begin{figure*}
\centerline{ 
\def\arraystretch{1.0}
\setlength{\tabcolsep}{0.0pt}
\begin{tabular}{lcr}
  \includegraphics[scale=0.290,keepaspectratio]{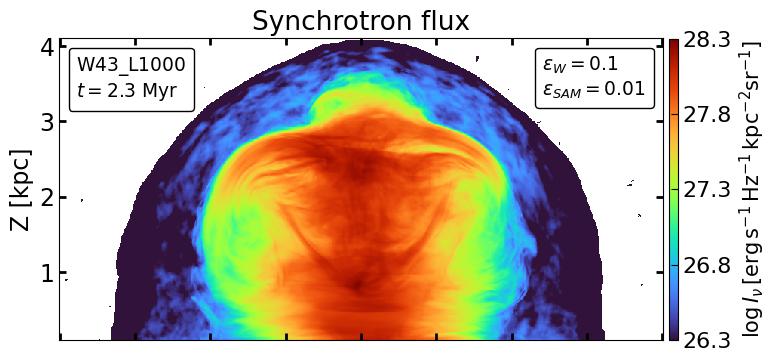}
    \includegraphics[scale=0.280,keepaspectratio]{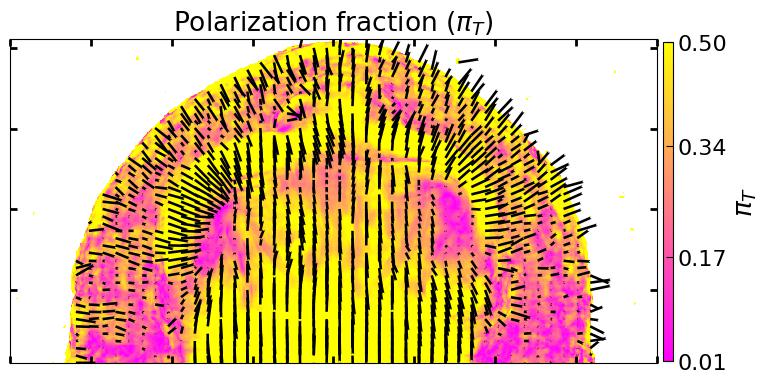}
    \includegraphics[scale=0.280,keepaspectratio]{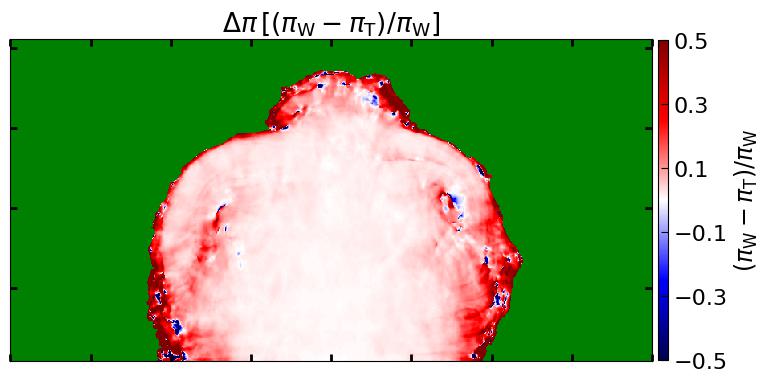} 
    \end{tabular}}
    \centerline{ 
\def\arraystretch{1.0}
\setlength{\tabcolsep}{0.0pt}
\begin{tabular}{lcr}
  \includegraphics[scale=0.294,keepaspectratio]{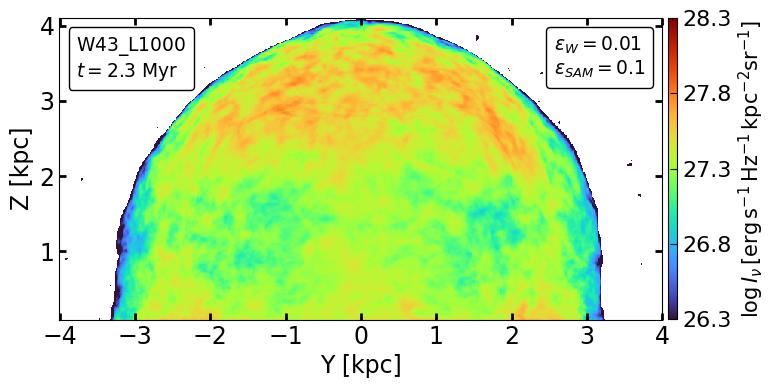}
    \includegraphics[scale=0.28,keepaspectratio]{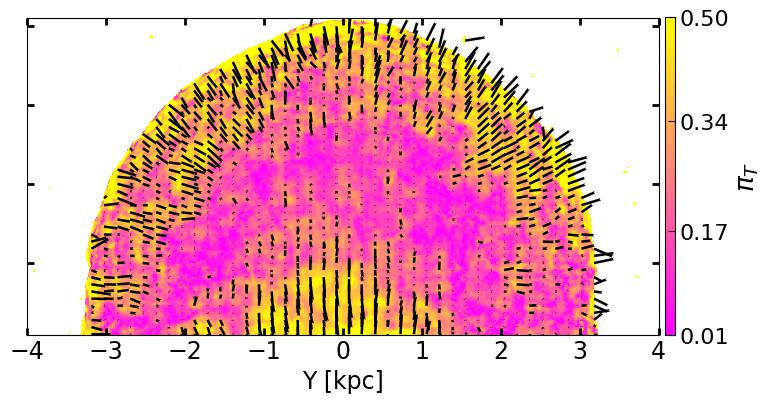}
    \includegraphics[scale=0.28,keepaspectratio]{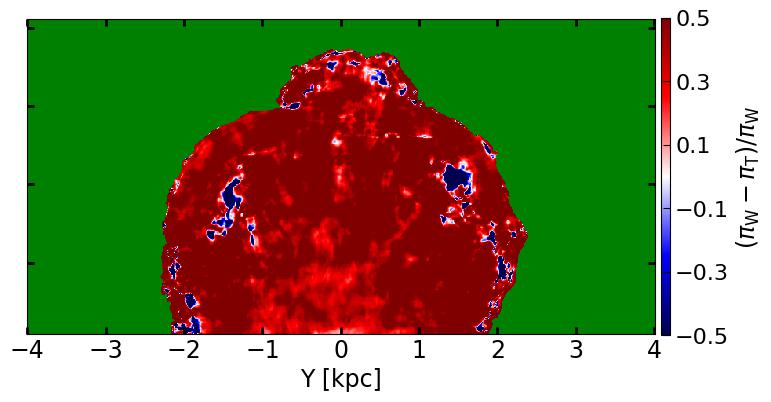} 
    \end{tabular}}

       \caption{Top: Logarithmic synchrotron flux (left), total polarization fraction (middle) and fractional change in polarization (right) at $\theta_I=90^\circ$ image plane for $\mathrm{W43\_L1000}$. The figures in the top row correspond to a stronger acceleration of electrons in the cocoon than the SAM ($\epsilon_{W}>\epsilon_{SAM}$), and conversely, in the bottom row.}
\label{fig:diff_epsilon}
\end{figure*}

\begin{figure*}
\centerline{ 
\def\arraystretch{1.0}
\setlength{\tabcolsep}{0.0pt}
\begin{tabular}{lcr}
\includegraphics[scale=0.58,keepaspectratio]{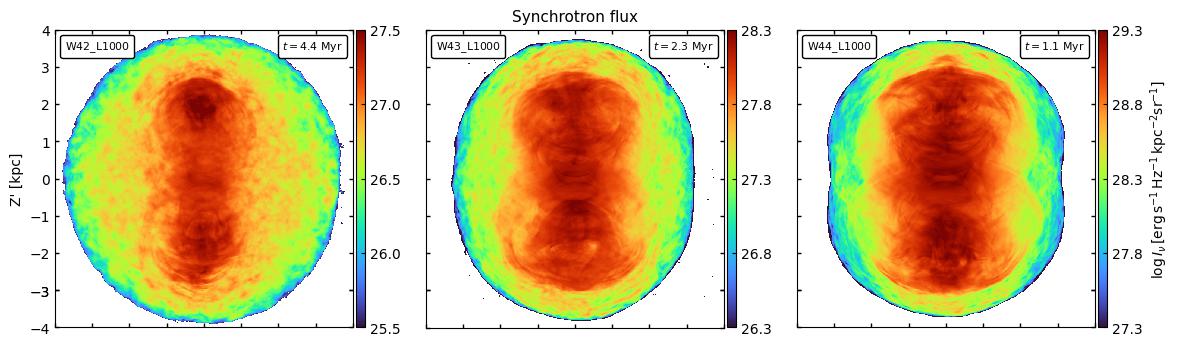}
    \end{tabular}}
    \centerline{ 
\def\arraystretch{1.0}
\setlength{\tabcolsep}{0.0pt}
\begin{tabular}{lcr}
\includegraphics[scale=0.58,keepaspectratio]{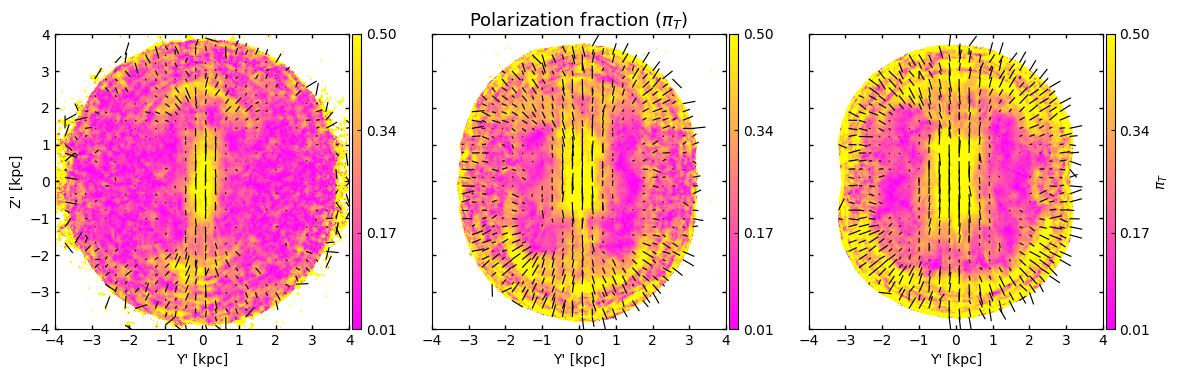}
    
    \end{tabular}}

       \caption{Logarithmic synchrotron flux (top), total polarization fraction (bottom) at $\theta_I=60^\circ$ image plane for simulations with different wind powers.}
\label{fig:diff_power_60deg}
\end{figure*}

\begin{figure*}
\centerline{ 
\def\arraystretch{1.0}
\setlength{\tabcolsep}{0.0pt}
\begin{tabular}{lcr}
  \includegraphics[scale=0.271,keepaspectratio]{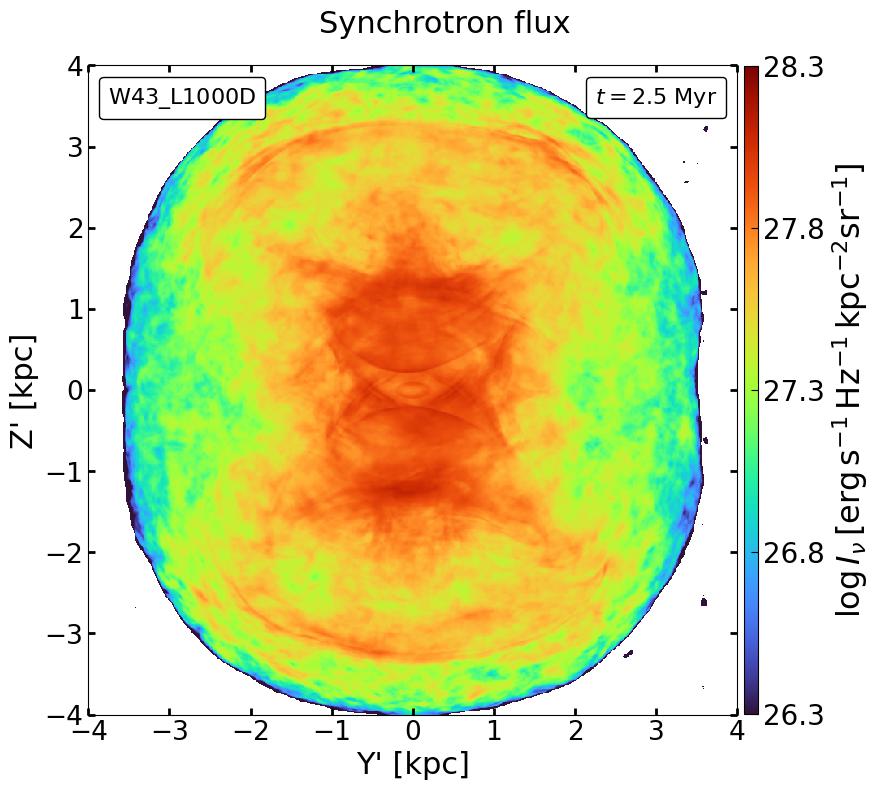}
   \includegraphics[scale=0.265,keepaspectratio]{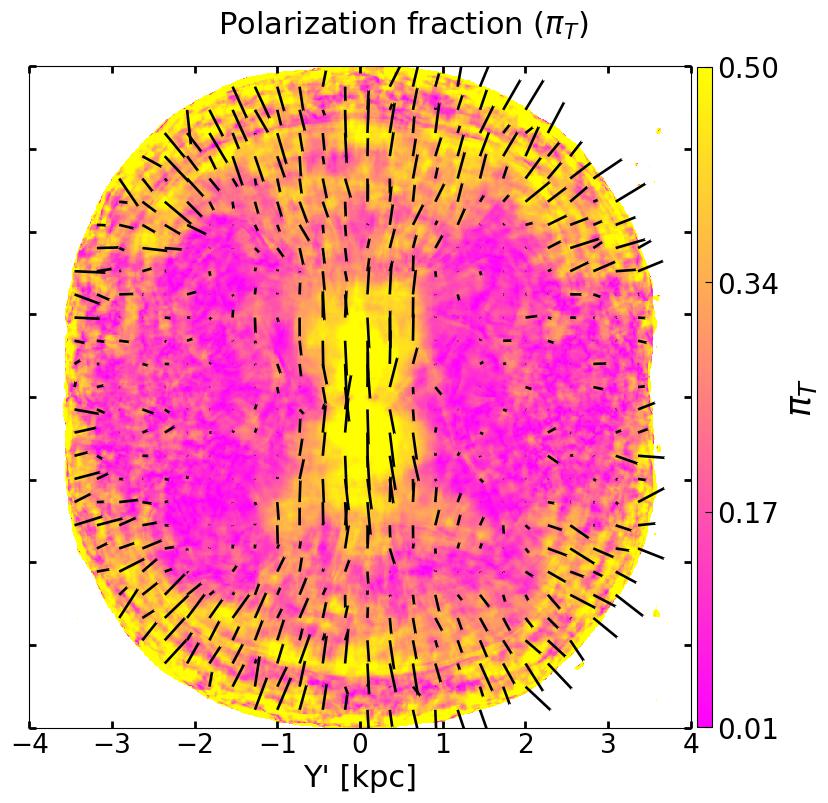}
    
    \end{tabular}}
       \caption{Logarithmic synchrotron flux  (left) and total polarization fraction (right) for $\mathrm{W43\_L1000D}$ at $\theta_I=60^\circ$ image plane.}
\label{fig:emiss_n1_60deg}
\end{figure*}

In this section, we discuss the synchrotron emission and polarization from the wind simulations listed in Table~\ref{tab:sim_table}. We present maps illustrating the logarithmic flux, polarization fractions ($\pi_T$ and $\pi_W$), and the fractional change in polarization ($(\pi_W - \pi_T)/\pi_W$) for winds with different powers at $\theta_I=90^\circ$ (angle from the +$Z$-axis) and $\phi_I=0$ (measured counter-clockwise from the X-Z plane) in Fig.~\ref{fig:diff_power_90deg}. The polarization vectors are overlaid on the polarization map with black arrows. Similar to Paper I, $\pi_T$ maps correspond to the total polarization obtained from the cocoon + SAM, and $\pi_W$ shows the polarization from the cocoon only. The conditions for identifying the regions inside the forward shock and SAM are mentioned in Appendix~\ref{forward_shock}. The fractional change in polarization due to the contribution of the SAM along the LOS is shown using $\Delta \pi = (\pi_W - \pi_T)/\pi_W$, i.e., $\Delta\pi = 1$ implies very strong depolarization due to the SAM, and $0$ means no depolarization. Fig.~\ref{fig:emiss_n2_90deg} showcase maps from dense ($\mathrm{W43\_L1000D}$) and highly dense ($\mathrm{W43\_L1000HD}$) winds. Additionally, in Fig.~\ref{fig:emiss_n2_th_90deg}, maps for a narrow light and dense wind ($\mathrm{W43\_L1000N}$ and $\mathrm{W43\_L1000HDN}$) are presented. In the paper, we assume $\epsilon_{W} \, \mathrm{and}\, \epsilon_{SAM}=0.1$, implying that the electrons carry exactly 10\% of the fluid energy density in the cocoon and SAM for almost all the cases. However, the acceleration efficiencies for non-thermal electrons in the cocoon and SAM can vary, which is explored in Fig.~\ref{fig:diff_epsilon}. Figs.~\ref{fig:diff_power_60deg} and~\ref{fig:emiss_n1_60deg} showcase maps depicting winds with different powers and the dense wind case, respectively, for an image plane at $\theta_I=60^\circ$. Below, we discuss the common characteristics of synchrotron emission and polarization, focusing on maps from the light winds at $\theta_I=90^\circ$ illustrated in Fig.~\ref{fig:diff_power_90deg}. How different physical parameters (wind power ($P_w$), density contrast ($\eta_w$), opening angle ($\theta_w$), viewing orientation of the observer ($\theta_I$), etc.) can affect these features in the winds is discussed in the subsequent section (Sec.~\ref{sec:synch_pol_vary}). The depolarization effect from the turbulent field in the SAM is discussed in Sec.~\ref{sec:depol_sam}.

\subsubsection{General characteristics of emission morphology and polarization}
\label{sec:synch}
\textbf{\underline{Morphology of emission:}}~

As can be seen from Fig.~\ref{fig:diff_power_90deg}, the general characteristics of wind emission at $\theta_I=90^\circ$ are diffuse spherical extended emission with a bright cylindrical structure that is capped by a horizontal bright band. However, the shapes of these substructures can vary depending on different parameters and also the observer's orientation, which we discuss in subsequent sections. The high emission in the central parts originates from the high-pressure zones situated between the lateral streams. Notably, nearly horizontal arcs of high emission are visible near the cocoon head (refer to H1, H2, and H3). Such a feature in the light winds arises due to elevated pressure and high magnetic fields, particularly the $B_y$ component of the wind's magnetic field in the $Y-Z$ planes\footnote{It is important to note that only the synchrotron emission resulting from the field components perpendicular to the line of sight contributes to the total emission at the designated image plane.}. 


Additionally, shearing of the fields (particularly $B_y$ and $B_z$) can also cause bright filamentary structures (F1 and F2) in the regions between the lateral streams. The vertical filaments are primarily influenced by the poloidal component (i.e., $B_z$) of the magnetic field, as can be seen from the middle row in Fig.~\ref{fig:B_xyz_43}. The lateral regions, characterized by backflows and eddies (refer to Fig.~\ref{fig:wind_structure}), depict low pressure, consequently exhibiting weak emission.

\textbf{\underline{Distribution of polarization:}}~
The polarization distribution ($\pi_T$) shows high polarization primarily from the forward shock, the central regions of the wind, and the horizontal arc, which coincides with a bright band in the synchrotron maps. High polarization at the forward shock arises from the compression of the external magnetic fields parallel to the shock. This outcome is similar to the findings observed for jets in Paper I. The polarization vectors align perpendicularly to the shock surface, indicating the compression of magnetic fields parallel to it. Within the SAM, there is a gradual decay of turbulent fields, resulting in low polarization levels due to the small-scale coherence of the fields. Regions above and below the high-polarized bright horizontal arc also display low-polarization values. This is due to turbulent fields in the SAM, and backflows in regions between lateral streams respectively in upper and lower zones. The poloidal fields between the lateral streams lead to low polarization in the cocoon, which is caused by the vector cancellation of orthogonally polarized components along LOS.

In $\pi_W$ (also $\pi_T$) maps, there is a prevalence of high polarization, attributed to the injected toroidal magnetic field within the wind, as indicated by the vertical polarization vectors. However, certain regions near the cocoon head and lateral sides exhibit low polarization values, resulting from the generation of poloidal fields. The poloidal fields are introduced due to velocity shears caused by backflows in these regions (see Fig.~\ref{fig:wind_structure}). Additionally, the mixing with the surrounding SAM, facilitated by eddies in the backflows, introduces further poloidal fields, which are subsequently amplified by turbulent motions within the cocoon (see $B_z$ map in Fig.~\ref{fig:B_xyz_43}). Such shearing of the fields near the contact surface leads to high polarization, particularly due to the small integration length in these regions. Consequently, the polarization vectors align perpendicularly to the projected surface of the cocoon in $\pi_W$ maps. 

\subsubsection{Effect of different parameters on the observable features}
\label{sec:synch_pol_vary}
\begin{itemize}
    \item \textbf{Dependence on the wind power:}
The emission maps in Fig.~\ref{fig:diff_power_90deg} indicate that the bright region in the central parts gets wider with increasing wind power. This, as shown in Fig.~\ref{fig:rho_prs_tr1}, is attributed to the weak collimation of the lateral streams, resulting in wide high-pressure zones above the Mach disk for the high-power winds. As a result, the bright horizontal arc is more extended as the wind power is increased. These arcs near the cocoon's head display a double-layered appearance for light low-power winds here, which is clearly visible for W42 and W43 and is less prominent for W44. The upper arc is formed due to the shearing of the $B_y$ component near the $Y-Z$ midplane, resulting from inward-directed backflows. This is evident from the evolution of the $B_y$ field displayed in the top row of Fig.~\ref{fig:B_xyz_43}, where fields are sheared to form an elongated, nearly horizontal structure. The lower arc is a consequence of the $B_y$ component of the toroidal magnetic field of the wind, coinciding with high-pressure zones at the top edges of the lateral streams. We confirm that such features may not be as distinctly observable from other lines of sight (e.g., in the $X-Z$ image plane), as it depends on the instantaneous flows in the wind cocoon. As indicated in Table~\ref{tab:sim_table}, the same magnetization value of 0.1 across all winds results in relatively weak magnetic fields for low-power winds in comparison to high-power ones. Consequently, as the cocoon expands, the local magnetic field gradually decreases, reaching a level comparable to that in the SAM (Fig.~\ref{fig:dist_B}). This results in nearly uniform emissions from the W42 wind. Conversely, high-power wind cocoons exhibit stronger magnetic fields than the SAM. Consequently, the emissions from the central parts of such cocoons are considerably higher compared to those emanating solely from the SAM.

From the polarization maps ($\pi_T$) in Fig.~\ref{fig:diff_power_90deg}, it can be seen that the high polarization at the forward shock is particularly evident in the W43 and W44 cases, but not in W42 due to weak forward shock (see Fig.~\ref{fig:rho_prs_tr1}). The polarization map in W42 shows larger areas of low polarization when compared to cases with higher power. This occurs as the SAM in W42 is broader, which enhances the decay and disorder of the magnetic fields within it. Consequently, the bright arc near the cocoon head is notably visible in the W42 case.

 \item \textbf{Dependence on the density of the wind:} 
 In order to compare how the wind density contrast can affect the observable features, we focus on the light, dense and highly dense wind cases with power $10^{43}~\ergs$. The maps from the light wind case are displayed in the middle column in Fig.~\ref{fig:diff_power_90deg}. The emission and polarization maps for the dense and highly dense winds at $\theta_I=90^\circ$ image plane are illustrated in Fig.~\ref{fig:emiss_n2_90deg}. It can be seen that the bright and highly polarized arc at the cocoon's head appears more extended for the dense winds when compared to the light wind case. This occurs due to more stability of the dense winds (as discussed in Sec.~\ref{sec:dynamics}) when compared to the light winds. Also, the absence of a double-layered structure in the horizontal bright band, as observed in lighter wind cases, is evident here. This absence, as illustrated in the bottom row of Fig.~\ref{fig:B_xyz_43}, is attributed to the lack of enhancement in the $B_y$ component within the $Y-Z$ midplane, a phenomenon resulting from inward-directed backflows in lighter wind cases. Consequently, the $B_y$ component of the toroidal field, coinciding with high-pressure zones at the top edges of the lateral streams (as depicted in the $B_x$ map in the bottom row), contributes to the formation of the bright horizontal arc when integrated along the line of sight in this scenario. Some emission can be seen between the Mach disk and horizontal arc for dense wind, giving rise to a ``mushroom-shaped'' morphology. This, however, is not seen for the highly dense wind, as the whole regions above the Mach disk contribute to the wide bright band here. 

Also, the horizontal arcs are quite prominent in the dense wind cases when compared to the light wind case. It can be seen that the emission from the central parts of the light wind is almost uniform, making the arcs less prominent. Inward-directed backflows in the light winds introduce toroidal as well as poloidal components between the lateral streams (see the middle row in Fig.~\ref{fig:B_xyz_43}), leading to almost homogeneous emission from regions above the Mach disk, which however is not the case for the dense winds (bottom row in Fig.~\ref{fig:B_xyz_43}). Contrarily, backflows in the dense wind mainly occur in the lateral sides of the high-velocity streams, which also generate poloidal fields. This leads to low polarization values in these regions, as can be seen from the $\pi_W$ maps.

 \item \textbf{Dependence on the opening angle of the wind:}  
Fig.~\ref{fig:emiss_n2_th_90deg} shows the emission and polarization maps for narrow light and highly dense winds, respectively, observed at an angle of $\theta_I=90^\circ$. The collimation of lateral streams in the light wind case leads to bright knots (labeled as N1). Both cases exhibit intense compact emissions labeled as H6 and N2, stemming from high-pressure zones near the cocoon head, which also coincide with high-polarization values. The highly dense wind case displays a horizontal bright and highly polarized arc from regions above the Mach disk (see H7 labeled region). However, in this instance, the width of the arc is relatively narrower, as the regions above are more constrained, resulting in a bright compact spot (N2) ahead. 
 
\item \textbf{Different particle acceleration scenarios:}
 In Fig.~\ref{fig:diff_epsilon}, the emission and polarization maps correspond to varying values of $\epsilon$ for the cocoon and SAM, i.e. implying different acceleration efficiency of the electrons in these regions. The top row shows the maps for higher $\epsilon_W$ (0.1) than $\epsilon_{SAM}$ (0.01); implying higher acceleration in the cocoon than SAM, and conversely in the bottom row. The electrons within the cocoon experience acceleration primarily at the Mach disk, while in the SAM, acceleration can be expected at the forward shock. When the efficiency of acceleration in the cocoon is notably high in the presence of magnetic fields \citep[for magnetically driven winds, see e.g.][]{fukumura_2015}, it results in stronger radio emissions originating from the cocoon compared to that from the SAM (as shown in the top row of Fig.~\ref{fig:diff_epsilon}). Conversely, if the SAM exhibits substantially higher acceleration efficiency \citep[for radiation-pressure-driven winds, see e.g.][]{nomura_2013}, the observable radio emission characteristics from the wind cocoon become less discernible (see bottom row). 

The polarization maps ($\pi_T$) indicate that when the wind is strong in accelerating electrons in the cocoon, most of the regions from the wind display high polarization. Conversely, the region exhibiting low polarization from the wind becomes more pronounced in the bottom row. This observation suggests that for radiation-driven winds, where the magnetic fields inside the winds may not be very strong, it is improbable to obtain significant radio polarization from the wind. However, high polarization can be observed from the forward shock region in SAM only if the emission is strong enough to be detected.

Moreover, it should be noted that the acceleration efficiency of the electrons at the forward shock can also vary in time. For example, as can be seen from density and pressure maps in Fig.~\ref{fig:rho_prs_tr1}, the forward shock in W42 has considerably weakened compared to the other cases, developing a double-layered structure. This weakening diminishes the impact of the outer shock on the compression of external magnetic fields, consequently lowering the acceleration efficiency of electrons in this region. This leads to low polarization values at the forward shock for W42, as can be seen from Fig.~\ref{fig:diff_power_90deg}. 

    
  \item \textbf{Dependence on the viewing angle:} At an inclined view of $\theta_I=60^\circ$ in Fig.~\ref{fig:diff_power_60deg} and~\ref{fig:emiss_n1_60deg} the emission from the central parts of wind appears ``hourglass-shaped''. The majority of this emission originates from regions situated above the Mach disk, with additional contributions coming from the areas surrounding the barrel zone, as illustrated at $\theta_I=90^\circ$ in Fig.~\ref{fig:diff_power_90deg}. Consequently, in light winds (see e.g. W42), these structures are narrower due to the strong collimation and relatively broader for the high-power light winds. The injected toroidal magnetic field results in high polarization from these regions in all the cases. In the surrounding areas with weak flux, low polarization is attributed to vector cancellation arising from orthogonally polarized components, along with contributions from the SAM. It should be noted that an inclined LOS, the toroidal field components are not uniformly perpendicular or parallel to the line of sight. This also add on to the low-polarized regions seen around the central spines. The distinctiveness of low polarization in W42 is emphasized due to its broader SAM, leading to rapid decay and disorder of the turbulent fields.
The broad annular rings at the top of the ``hourglass'' structure (particularly visible on the approaching side for W43 and W44) represent high-pressure annular regions above the edges of the lateral streams coinciding with high toroidal field components. Due to weak inward-directed backflows in the dense wind when compared to the light winds, any emission from the regions between the arcs is considerably weaker. Nonetheless, the bright outer arcs display high polarization in all cases. These arcs are distinctly visible for W42, whereas they seem merged with the high-polarization regions from the SAM in others.
\end{itemize}

\subsubsection{Depolarization effect of the SAM}
\label{sec:depol_sam}
\begin{itemize}

   \item \textbf{Dependence on the wind power:} 
  The fourth row in Fig.~\ref{fig:diff_power_90deg} depicts the fractional change in the polarization of the cocoon ($\Delta \pi$) after the contribution from the SAM. More reddening for the W42 clearly indicates that the presence of a turbulent field in the SAM leads to more lowering in the polarization from the cocoon for the low-power winds. This is because the magnetic fields in the SAM are quite comparable to the cocoon for the low-power winds, as can be seen from the PDFs of the magnetic fields in Fig.~\ref{fig:dist_B}. Contrarily, the high-power winds carry stronger magnetic fields than the SAM, and thus the depolarization effect is weaker. As found in Paper I, the magnetic fields in the cocoon and SAM decay with the expansion. When the fields in both regions become comparable, the depolarization from the SAM gets strengthened.
   
\item \textbf{Dependence on $\epsilon_W$:}
It can be seen from $\Delta \pi$ maps in Fig.~\ref{fig:diff_epsilon}, the depolarization from the SAM becomes prominent for the case with $\epsilon_{SAM}=0.1$ and $\epsilon_W=0.01$. In the context of a radiation-driven wind, the non-thermal radio emission from the wind cocoon may not be very strong. Under such circumstances, the SAM's contribution takes precedence in the total non-thermal emission, as illustrated in the left panels of the figure. Consequently, the diminished polarization of the wind primarily results from the contribution of turbulent magnetic fields within the SAM.


\end{itemize}

\begin{figure*}
\centerline{ 
\def\arraystretch{1.0}
\setlength{\tabcolsep}{0.0pt}
\begin{tabular}{lcr}
\includegraphics[scale=0.7,keepaspectratio]{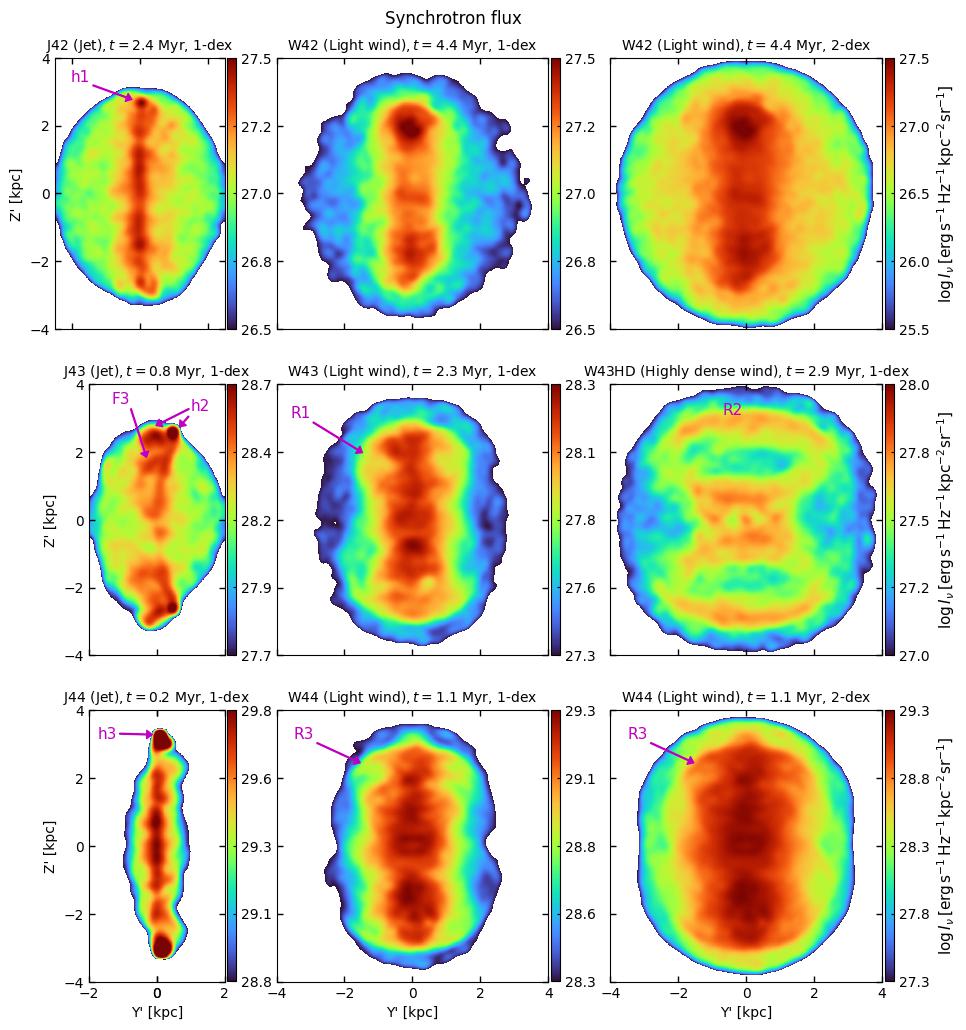} 
    \end{tabular}}
       \caption{Logarithmic synchrotron flux from jet simulations in Paper I and wind simulations in this study, respectively, is presented at the $\theta_I=60^\circ$ image plane. The color bar range is maintained at one order of magnitude in logarithm in all figures except for the top and bottom right ones, where a two-orders-of-flux range is used for comparison. Regions with flux values lower than the minimum in the color bar are removed, and the flux is convolved with a Gaussian filter with an FWHM of 240~pc. The hotspots in J42, J43 and J44 are indicated as h1, h2  (`multiple') and h3, respectively. F3 represents a bright filamentary-like structure in J43. R1, R2, and R3 represent the bright annular rings near the cocoon head in the winds.}
\label{fig:synch_jet_wind_100pc}
\end{figure*}

\begin{figure*}
\centerline{ 
\def\arraystretch{1.0}
\setlength{\tabcolsep}{0.0pt}
\begin{tabular}{lcr}
\includegraphics[scale=0.7,keepaspectratio]{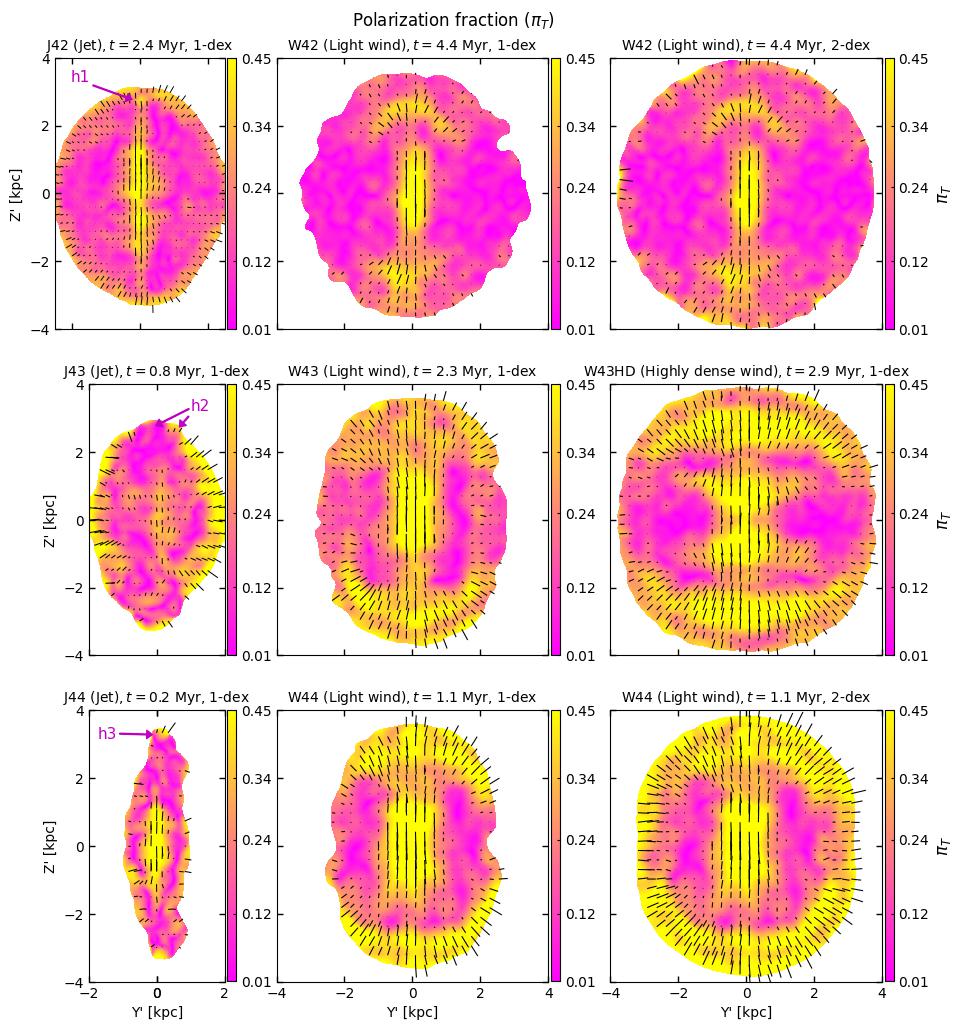} 
    \end{tabular}}

       \caption{Convolved total polarization fraction ($\pi_T$) corresponding to the emission maps displayed in Fig.~\ref{fig:synch_jet_wind_100pc}. The Gaussian filter used for convolution has a FWHM of 240~pc. }
\label{fig:pol_jet_wind_100pc}
\end{figure*}

\begin{figure*}
\centerline{ 
\def\arraystretch{1.0}
\setlength{\tabcolsep}{0.0pt}
\begin{tabular}{lcr}
  \includegraphics[scale=0.7,keepaspectratio]{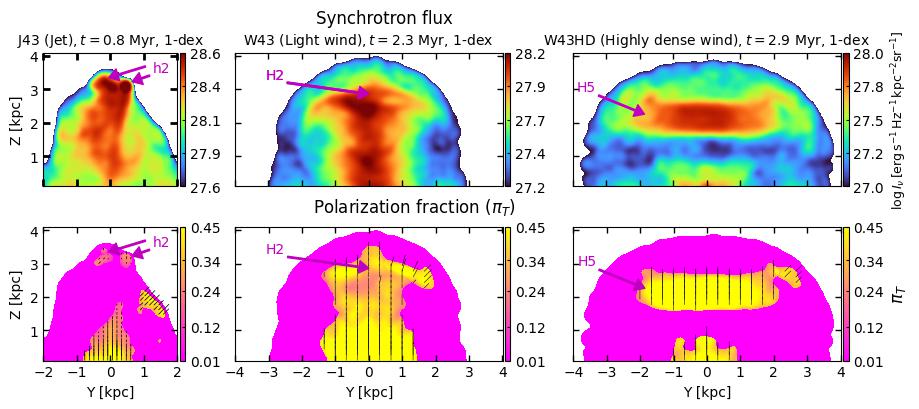}
    \end{tabular}}
       \caption{Logarithmic synchrotron flux (top) and total polarization fraction (bottom) for J43 and W43 at $\theta_I=90^\circ$ image plane. Both emission and polarization maps correspond to a one-order-of-magnitude flux range and are convolved with a Gaussian filter with an FWHM of 240~pc. The multiple hotspots in J43 are shown as h2, and the bright horizontal arcs in W43(HD) are presented as H2 and H5.}
\label{fig:wind_jet_90deg}
\end{figure*}

\begin{table*}
     \centering
     \caption{Selection criteria to distinguish between jets and winds. These traits are discussed in detail in Sec.~\ref{sec:jet_wind_compare}.}
     \label{tab:jet_wind_criteria}
     \begin{threeparttable}
	
     \begin{tabular}{ |L{3cm} | L{2cm} | L{3cm} | L{3.5cm} | c | }
      \hline
        Characteristic features & Jets &  Winds &  Comment & Criteria \\ \hline
          Regions of highest flux & Spine and hotspots(s)  &  Similar, but broader spine and arcs & Resemblance for some models with low resolution & Maybe \\ \hline
          Width of bright emission & Narrow & Comparatively broader & Can be comparable for low resolution & Maybe\\  \hline
         High polarization & Localized: hotspots, spine & Wide arcs and spine &  Persist with lowering resolution, but low-values dominate & Yes \\  \hline
         Depolarization in cocoon & More &  Less & Same as above  &  Yes \\  \hline
         Collimation & Mostly collimated, cocoon width varies with power & Dense winds are broader; light winds (especially low power) can undergo some collimation & Low-power jets have broader cocoons, while low-power winds, due to collimation, have narrower cocoons than their high-power counterparts & Maybe \\ \hline
         
     \end{tabular}
     \end{threeparttable}
	
 \end{table*}

\section{Comparing synchrotron and polarization characteristics from jets and winds}
\label{sec:jet_wind_compare}

The primary motivation of this section is to identify some noticeable observational features for the non-thermal synchrotron emission and polarization to distinguish between the jets and winds. Emission and polarization maps depicting jets and winds with varying powers and viewing angles are presented in Figs.~\ref{fig:synch_jet_wind_100pc}-\ref{fig:wind_jet_90deg} and Figs.~\ref{fig:synch_jet_wind_300pc}-\ref{fig:pol_jet_wind_300pc}. To compare with observed results, these maps are convolved with a Gaussian filter with a different standard deviation of 100~pc and 300~pc and truncated at twice these scales\footnote{This implies FWHM of 240 and 720~pc, and truncated at scales of 400 and 1200~pc, respectively.}. The corresponding FWHM gives an angular resolution of 125 and 375 milliarcsec, respectively, for sources at a redshift ($z$) = 0.1 (assuming $H_0= 67\, \mathrm{kms^{-1} Mpc^{-1}}$, and $\Omega_M=0.3$). The latter values are comparable to the resolution scales reached with previous studies \citep[see e.g.][]{jarvis_2019, jarvis_2021,pierce_2020} and can be achieved in the local universe \citep{baldi_2018}. For a larger redshift of 2, they correspond to a resolution of around 27.4 and 82~mas, which is typically not achieved by the studies looking at this scale \citep{alexanderoff_2016,rosario_2021}; although can be possible using high-resolution data \citep{njeri_2024}. In these figures, we mainly focus on one order in magnitude for the flux dynamic range in the fiducial maps presented here and also show maps for the two-order-of-magnitude flux range for W42 and W44. The polarization vectors are depicted as black arrows. These snapshots correspond to times when the jet or wind head approaches near the top of the $Z$ axis, and the forward shock is confined within the box. The correlation length of the external magnetic field is 1000~pc in all the simulations, except for J42, where, we ran the simulation with only a 500~pc coherence scale in Paper I.

\subsection{Characteristics in emission morphology}
The winds and jets display some common traits as well as some differences in emission morphology, which are listed below:
\begin{itemize}
    \item In the emission morphologies, it is evident that bright features emerge from the central regions in both jets and winds, which are surrounded by weak extended emission (see Figs.~\ref{fig:synch_jet_wind_100pc} and~\ref{fig:wind_jet_90deg}). A jet has a systematically narrow spine with defined hotspot, or hotspot complexes for kink-unstable jets (see h1, h2, and h3). Winds, on the other hand, display broader spines or ``hourglass-shaped'' bright emission. The bright annular rings/arcs observed at an inclined view of $60^\circ$ (R1, R2, and R3) manifest into bright horizontal arcs near the edges of the cocoon at $90^\circ$ (H2 and H5). \textit{Such rings/arcs are not seen for jets and are a particular feature in the winds.}
    
    \item Winds exhibit a diverse range of morphologies depending on their power and density. Light and low-power winds tend to become collimated, unlike high-power winds. \textit{As a result, the presence of a hotspot may not reliably distinguish jets from winds, as pronounced collimation in low-power winds can produce similar features} (as observed for W42 in Fig.~\ref{fig:synch_jet_wind_100pc}). We have confirmed that such strong collimation and hotspot-like regions do not form in highly dense ($\eta_w=4$) wide-angled winds with a power of $10^{42}~\ergs$. Additionally, the emission closely resembles that of the dense wind case illustrated in Fig.~\ref{fig:emiss_n1_60deg}. This suggests that dense winds are more stable and thus, exhibit prominent bright annular rings (and bright distinguished horizontal arcs at $\theta_I=90^\circ$, as inferred from Fig.~\ref{fig:wind_jet_90deg}) compared to lighter winds. However, it is worth noting that the hotspot-like features can still form in dense narrow-angled winds, as depicted in Fig.~\ref{fig:emiss_n2_th_90deg}.

  \item \textit{Effect of dynamic range:} To illustrate the impact of dynamic range, we present the emission maps for W42 and W44 using a two-orders-of-magnitude flux range in Fig.~\ref{fig:synch_jet_wind_100pc} (see top and bottom right plots). Notably, the extended emission is narrower for jets than for winds. It can be observed that the emission from regions near the forward shock in the SAM is lower compared to that from the central parts of the wind. Consequently, while the central bright regions are unlikely to be missed when observations are limited to one order of magnitude in dynamic range, outer regions, which are part of the SAM, can be overlooked.

  \item \textit{Effect of resolution:}
  The extent of bright emission from the central parts varies in both jets and winds with resolution. To facilitate comparison, we calculated the average width of the regions surrounding the central vertical axis, encompassing 80\% of the maximum flux at the given $Z'$ value, within a range of $Z'=\pm 1$~kpc in Fig.~\ref{fig:synch_jet_wind_100pc}. At high resolutions, these widths measure 420~pc and 310~pc for J42 and J44, respectively, and approximately 1.2~kpc and 1.3~kpc for W42 and W44.  In low-resolution maps shown in Fig.~\ref{fig:synch_jet_wind_300pc}, the widths of the bright spines have increased to quite comparable values of 920~pc and 1.2~kpc for J42 and W42, while for J44 and W44, these are 630~pc and 1.36~kpc, respectively. Furthermore, it is apparent that at low resolution, several features, such as the spine, hotspots, filaments, and arcs, have become less prominent. This leads to some resemblance in appearance between jets and some wind models. For example, in light winds, bright ``hourglass-shaped'' regions appear close to the spine of the jets. Interestingly, bright regions in W42 closely match the emission morphology from the jet in J44. Although a hotspot in W42 might resemble that of a jet, it is important to recognize the absence of a bright, thin spine typically associated with jets, which becomes evident at higher resolutions. \textit{Since the emission morphology is often used to distinguish between jets and winds, our study shows that some features can become blended when poor resolution is used.} This underscores the importance of higher-resolution radio observations; however, noting that ``only'' utilizing the very highest resolution images may resolve away important diffuse structures that also contain important diagnostic information in radio morphology and polarization. 

\end{itemize}

\subsection{Distribution of polarization}
Below, we list some common characteristics and differences in the polarization structure from the jets and winds.
\begin{itemize}
    \item Some notable features can be clearly seen from the maps depicted in Figs.~\ref{fig:pol_jet_wind_100pc} and~\ref{fig:wind_jet_90deg}. \textit{High polarization is mainly detected along the central spine and hotspots/arcs, which are localized in jets and are more widespread in the winds.} Additionally, high polarization is observed along the filament (F3) within the jet, although it is reduced due to smoothing.

    \item In jets, particularly those with a higher power, the hotspots exhibit noticeable high-polarization values, typically found near the upper edges of the cocoon. These hotspot(s) exhibit compactness, resulting in polarization originating from them being concentrated at specific locations (e.g. refer to h1, h2 h3). However, in winds, high polarization mainly occurs from the central parts and top edges of the cocoon (refer to R1, R2, R3, H2, and H5 from emission maps). These arcs are surrounded by comparatively low-polarization values (e.g. see maps for W42 and W43 in Fig.~\ref{fig:pol_jet_wind_100pc}). The lateral edges of the maps can also display high polarization values, as can be seen particularly in the middle row left panel in Fig.~\ref{fig:pol_jet_wind_100pc}.

    

\item In Fig.~\ref{fig:pol_jet_wind_100pc}, both winds and jets display low polarization values in their extended regions, which mainly arises because of the vector cancellation of orthogonally polarized components, and also some contributions from the SAM. \textit{It is clear that in the jets' cocoon, where stronger poloidal fields are generated due to pronounced backflows, experiences more internal depolarization compared to winds} (e.g. see $\pi_J$ maps in Fig.~9 of Paper I and $\pi_W$ in Figs.~\ref{fig:diff_power_90deg}-\ref{fig:emiss_n2_90deg}). The higher prevalence of low polarization in W42 is linked to its wider SAM and relatively narrower cocoon, as depicted in Fig.~\ref{fig:rho_prs_tr1}, resulting in more decay of turbulent fields.

 \item \textit{Effect of dynamic range:} The top and bottom right plots for total polarization (see Fig.~\ref{fig:pol_jet_wind_100pc}) indicate that a two-order-of-magnitude flux range enables the observation of polarized emission from the outer parts of the SAM. This remains invisible with a lower dynamic range, as shown in the plots on the left side. These regions may exhibit either high or low polarization values in different wind power simulations. In the case of W42, the weak strength of the forward shock leads to low polarization, whereas relatively higher polarization is seen in W44. The lateral regions in W44 (which corresponds to SAM) also display high-polarization values, similar to those seen for the jets (see left panels), which were not observed at the lower dynamic range. At $\theta_I=90^\circ$, high values like these are not observed in 1-dex range maps (see Fig.~\ref{fig:wind_jet_90deg}). However, we confirm that they are seen, primarily near the bow shock's head, in 2-dex range maps.

   \item \textit{Effect of resolution:}
In Fig.~\ref{fig:pol_jet_wind_300pc}, it is evident that decreasing the resolution has not altered the qualitative distribution of polarization characteristics. However, low-polarization values have become more prevalent. Polarization values near the spine and hotspots in jets appear lower compared to the high-resolution maps. In winds, the highly polarized arcs near the cocoon head seem to blend with regions from the SAM, and the resulting values are diminished after smoothing. \textit{This suggests that polarization can be an important factor in distinguishing between the jets and winds.}

\end{itemize}

\section{Summary and Discussion}
\label{sec:discussion}

AGN-driven jets and winds are believed to be launched by distinct mechanisms \citep{konigl_2006}. Numerous simulations have been performed for both \citep[see e.g.][]{proga_2004, perucho_2007, fukumura_2015, perucho_2022}, shedding light on various aspects relevant to real systems. It is found that the evolution and stability of jets are influenced by factors such as Mach number, Lorentz factor, and density contrast with respect to the ambient medium \citep{bodo_1994,perucho_2005}. Conversely, the understanding of winds remains less clear, with their origin being a subject of ongoing debate. It is plausible that a single mechanism may not suffice to explain the observed ultrafast outflows (UFOs) in quasars, suggesting the contribution of multiple processes \citep{wang_2022}. This underscores the necessity for incorporating more sophisticated physics and a deep understanding of physical conditions in AGNs to comprehensively account for the winds.

Observations have identified clear indications of AGN-driven winds, whether through direct or indirect means. For instance, an inverse relationship between the radio loudness and the column density of ionized wind is found in radio-loud AGN by \citet{mehdipour_2019}. This connection suggests a shared magnetic mechanism responsible for initiating both wind and jet formations \citep[see e.g.][]{king_2013}. Excess radio emissions within several quasars also appear to be predominantly influenced by the presence of either winds or small-scale jets \citep{nadia_2014,jarvis_2019,petley_2022,andonie_2022}. 

Theoretical simulations can help establish a connection with observational findings, bridging the gap in our understanding of AGN-driven jets and winds. In our work (Paper I and this paper), we have attempted to gain insights into the evolution of jets and winds within a medium enriched with turbulent magnetic fields. We have examined the behavior of these at different energies, all of which are initiated with a toroidal magnetic field in our studies. Our analysis reveals that the evolution of jets and winds follows different behaviors, leading to observable distinctions in their features, which we summarize in the following section.

\subsection{Dynamical and observational perspectives on jets and winds}

As indicated in Sec.~\ref{sec:dynamics} of this study and Sec.~4.1 in Paper I, the evolution of jets and winds exhibits distinct evolution characteristics. Jets, characterized by high beaming and relativistic speeds, advance more rapidly in their surroundings compared to winds. In contrast, winds exhibit mild relativistic speeds with a wide-angle spread, providing a larger surface area for interaction with the surrounding medium. This wide-angle nature results in instabilities at the periphery of the barrel zone, particularly evident in light winds, leading to lateral streams from the sides of the barrel (see Fig.~\ref{fig:wind_structure}). These streams influence wind evolution by inducing backflows ahead of the Mach disk and contributing to mixing with the SAM. Dense winds, in contrast, demonstrate stability and propagate without being much affected by their surroundings. The emission and polarization features from the winds with varying parameters (such as power, density, and opening angle) are discussed in detail in Sec.~\ref{sec:synch_pol}.

We have analyzed polarization and emission from jets and winds with different models and compared them in Sec.~\ref{sec:jet_wind_compare}. The insights from the observable distinguishing traits between jets and winds are outlined in Table~\ref{tab:jet_wind_criteria}. We find that several characteristic features, such as hotspots and spines, may not be good indicators for poorly resolved sources, highlighting the need for highly resolved observations. The polarization distribution appears to be a reliable indicator for distinguishing between jets and winds, as it is not significantly affected by resolution. Jets exhibit compact, bright emissions from narrow spines and hotspots, aligning with high polarization values. Winds display similar widespread features, with the polarized spine nearly hourglass-shaped and capped with polarized bright arcs. Additionally, our studies suggest that depolarization, due to poloidal fields generated in the cocoon, is more dominant in jets compared to winds. It should be noted that both jets and winds are initially launched with a toroidal magnetic field in our study; nevertheless, a poloidal component at the base region is expected to exist, giving rise to helical magnetic fields. Consequently, the amplification of the poloidal field is likely to be much more pronounced in jets compared to winds. Contrarily, a weak poloidal component is seen in winds, with a dominance of toroidal field elsewhere in the cocoon. In several radio sources, such transverse or toroidal components at the base of the jets (with poloidal fields) are associated with AGN wind \citep{silpa_2021a,silpa_2021b}. The jets remain collimated over longer spatial scales, although the cocoons widen as jet power decreases. However, denser winds consistently remain broad. Lighter, low-power winds can become unstable at earlier stages, leading to collimation by the surroundings and resulting in narrower cocoons than those of their high-power counterparts.

\subsection{Limitations and future aspects}
In our studies (Paper I and this work), we have focused on the evolution of magnetized compact collimated jets and wide-angled winds in a hot halo carrying turbulent magnetic fields. However, in realistic scenarios, such compact jets/winds are expected to interact with the multiphase ISM of the host galaxy. This interaction, in turn, may also influence their synchrotron and polarization characteristics, which we reserve for future exploration. In addition to investigating the morphology of emission and polarization, simulations can also provide valuable quantitative parameters, such as spectra originating from the acceleration of non-thermal electrons within the jets and winds. Particle acceleration processes are likely to differ between jets and winds. For example, significant acceleration in jets may occur along the spine and at the hotspot, while in winds, the Mach disk can accelerate particles. Furthermore, differences in the strength of forward shocks and SAM properties in both jets and winds may have varying effects on particle spectra originating from the SAM. This aspect is reserved for future research, wherein non-thermal electrons \citep[Lagrangian Particle in \textsc{pluto}:][]{vaidya_2018} will be introduced into the simulation domain and their evolution will be followed. Theoretical results can subsequently be juxtaposed with direct predictions from recent observational missions like LOw-Frequency ARray (LOFAR)\footnote{https://lofar-surveys.org/}, eMERLIN\footnote{https://www.e-merlin.ac.uk/}, Event Horizon Telescope (EHT)\footnote{https://eventhorizontelescope.org/}, and Very Large Array (VLA)\footnote{http://www.vla.nrao.edu/}, thereby facilitating a more clear understanding of jets and winds.

\section*{Acknowledgements}
 We express our gratitude to the Accordo Quadro INAF-CINECA 2017 and `Pegasus' HPC facility at IUCAA, Pune, India, for high-performance computing resources. We also acknowledge the assistance of the INDO-ITALIAN mobility grant (INT/Italy/P-37/2022(ER)(G)). GB and PR acknowledge support by Next Generation EU through  PRIN MUR 2022 (grant n. 2022C9TNNX) and by the INAF Theory Grant ``Multi-scale simulations of relativistic jets''. C.M.H acknowledges funding from a United Kingdom Research and Innovation grant (code: MR/V022830/1). MM expresses gratitude to Ankush Mandal for the valuable discussions.

\section*{DATA AVAILABILITY}
The simulations generated for this study will be shared upon a reasonable request to the corresponding authors.


\bibliographystyle{mnras}
\bibliography{manuscript} 

\appendix
\section{Identification of forward shock and SAM}
\label{forward_shock}
The SAM region in different simulations is identified using the pressure ($P$), density ($\rho$), and jet-tracer ($\mathrm{tr1}$) as:

\begin{gather}\label{eq:forward_shock}
 \textbf{W42:} ~(P>2 \times P_0\, \text{or} \, \text{tr1}>0\, \text{or} \,\rho > 0.007~\text{cm}^{-3})\, \&\text{\,tr1} \leq 10^{-3} \nonumber \\
  \textbf{W43(N):} ~(P>2 \times P_0\, \text{or}\, \text{tr1}>10^{-20}) \, \&\text{\,tr1} \leq 10^{-3}  \nonumber \\
   \textbf{W43D:} ~(P>1.6 \times P_0\, \text{or}\, \text{tr1}>10^{-20}) \, \&\text{\,tr1} \leq 10^{-3}  \nonumber \\
    \textbf{W43HD:} ~(P>1.5 \times P_0\, \text{or}\, \text{tr1}>10^{-20}) \, \&\text{\,tr1} \leq 10^{-1}  \nonumber \\
    \textbf{W43HDN:} ~(P>2 \times P_0\, \text{or}\, \text{tr1}>0) \, \&\text{\,tr1} \leq 10^{-4}  \nonumber \\
 \textbf{W44:} ~(P>4 \times P_0\,\text{or}\, \text{tr1}>10^{-20})\, \&\text{\,tr1} \leq 10^{-3}
\end{gather}

Here, $P_0$ is the initial pressure in the simulation domain. The conditions in the brackets select the regions inside the forward shock. The SAM is then identified by the wind tracer limit at the end, which is also used to identify the regions inside the wind cocoon.


\section{Evolution of magnetic fields in the light and highly dense winds}
\begin{figure*}
\centerline{ 
\def\arraystretch{1.0}
\setlength{\tabcolsep}{0.0pt}
\begin{tabular}{lcr}
\includegraphics[scale=0.67,keepaspectratio]{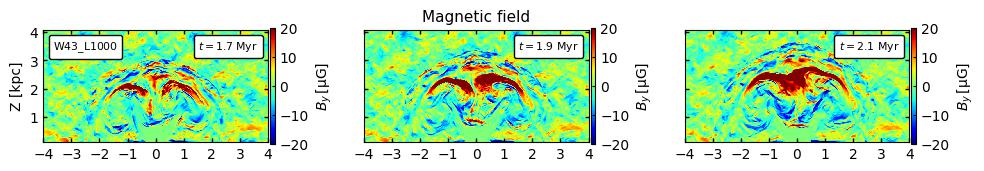} 
    \end{tabular}}
\centerline{ 
\def\arraystretch{1.0}
\setlength{\tabcolsep}{0.0pt}
\begin{tabular}{lcr}
\includegraphics[scale=0.67,keepaspectratio]{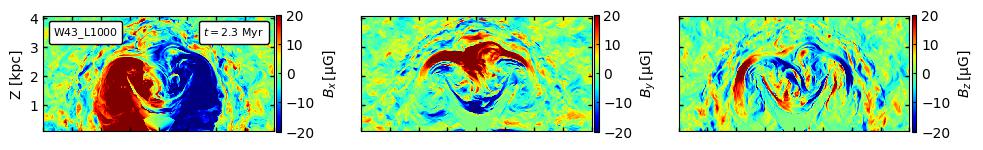} 
    \end{tabular}}
    \centerline{ 
\def\arraystretch{1.0}
\setlength{\tabcolsep}{0.0pt}
\begin{tabular}{lcr}
\includegraphics[scale=0.67,keepaspectratio]{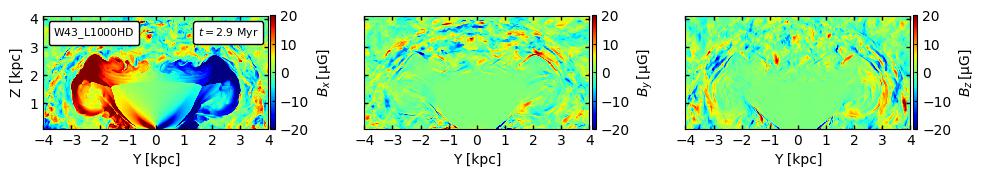} 
    \end{tabular}}
       \caption{Top: Time evolution of the $B_y$ component of the magnetic field in the $Y-Z$ midplane for light wind run of W43 from Table~\ref{tab:sim_table}. An elongated horizontal structure with high $B_y$ values forms near the cocoon head with time. This high-field component gives rise to the upper bright horizontal arc observed in the middle top panel of Fig.~\ref{fig:diff_power_90deg}. Middle and Bottom: $X, Y$ and $Z-$components of the magnetic field in the $Y-Z$ plane for light ($\mathrm{W43\_L1000}$, top row) and highly dense wind ($\mathrm{W43\_L1000HD}$; bottom row) simulations of W43. $B_y$ component of the toroidal field gives rise to a horizontal bright arc in the emission maps (as discussed in Sec.~\ref{sec:synch}) at $\theta_I=90^\circ,\phi_I=0^\circ$ image plane.}
\label{fig:B_xyz_43}
\end{figure*}

Fig.~\ref{fig:B_xyz_43} illustrates the time evolution of the $B_y$ component of the magnetic field for the light wind W43 in the top row. It can be seen that the inward-directed backflows cause shearing and amplification of $B_y$ fields near the cocoon's head. The substructures of fields from opposite sides can combine, forming an elongated amplified magnetic component. Such phenomenon gives rise to the upper bright horizontal arcs in the light wind cases, depicted in Fig.~\ref{fig:diff_power_90deg}.

 In the middle and bottom rows, we have shown the different magnetic field components in the $Y-Z$ midplane for the light and highly dense winds, respectively. It is clear that the evolution of wind differs between these cases, as elaborated in Sec.~\ref{sec:dynamics}, resulting in varying magnetic field structures. The $B_x$ component represents the injected toroidal field and exhibits enhanced values in both cases. As shown in the top row, the inward-directed backflows enhance the $B_y$ component of the field near the cocoon head in light wind. Moreover, the $B_z$ component is also introduced above the Mach disk. On the contrary, a highly dense wind does not exhibit a significant enhancement in the magnetic field ($B_y$ and $B_z$) above the Mach disk. The dominating toroidal component is notably higher in this case, particularly near the top edges of the lateral streams, which have partially emerged in this case.

\section{Jets vs winds: synchrotron emission and polarization}
\label{sec:jet_wind_images}

\begin{figure*}
\centerline{ 
\def\arraystretch{1.0}
\setlength{\tabcolsep}{0.0pt}
\begin{tabular}{lcr}
\includegraphics[scale=0.7,keepaspectratio]{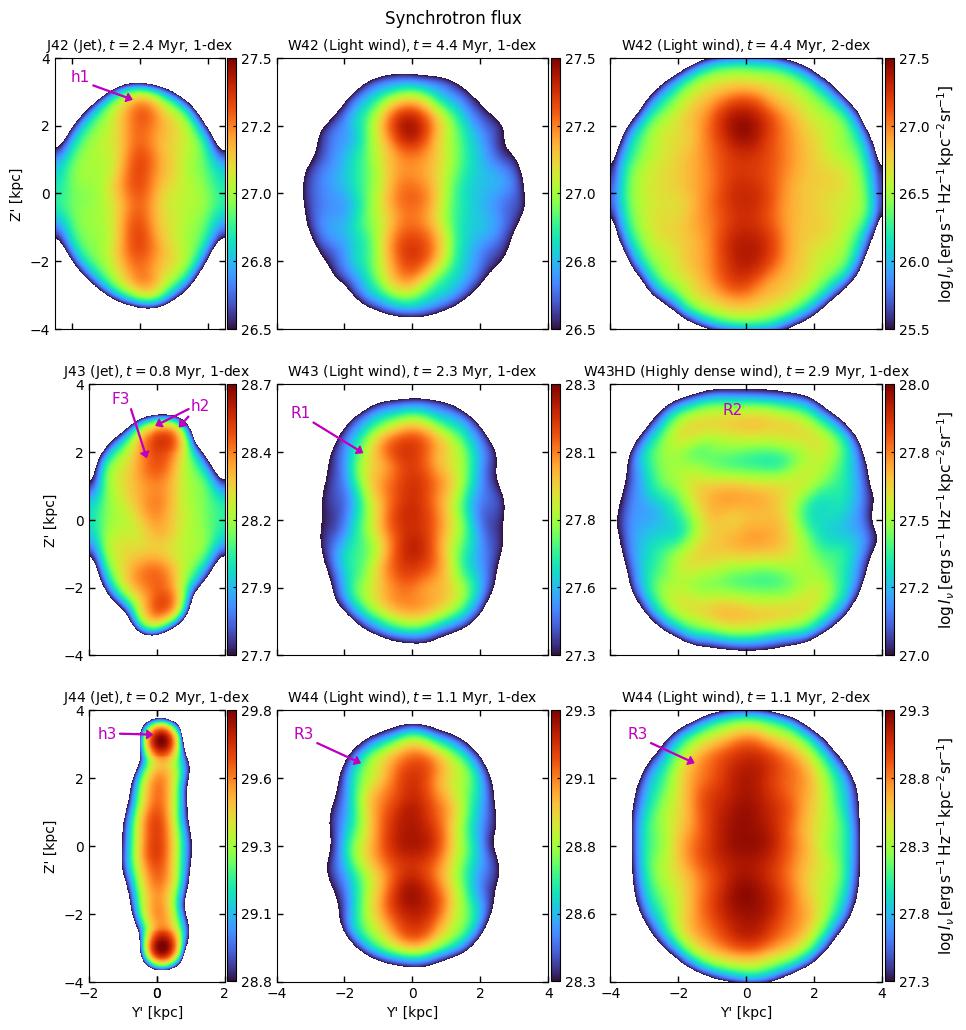} 
    \end{tabular}}
       \caption{Same as Fig.~\ref{fig:synch_jet_wind_100pc}, but convolved with a Gaussian filter with a FWHM of 720~pc.}
\label{fig:synch_jet_wind_300pc}
\end{figure*}

\begin{figure*}
\centerline{ 
\def\arraystretch{1.0}
\setlength{\tabcolsep}{0.0pt}
\begin{tabular}{lcr}
\includegraphics[scale=0.7,keepaspectratio]{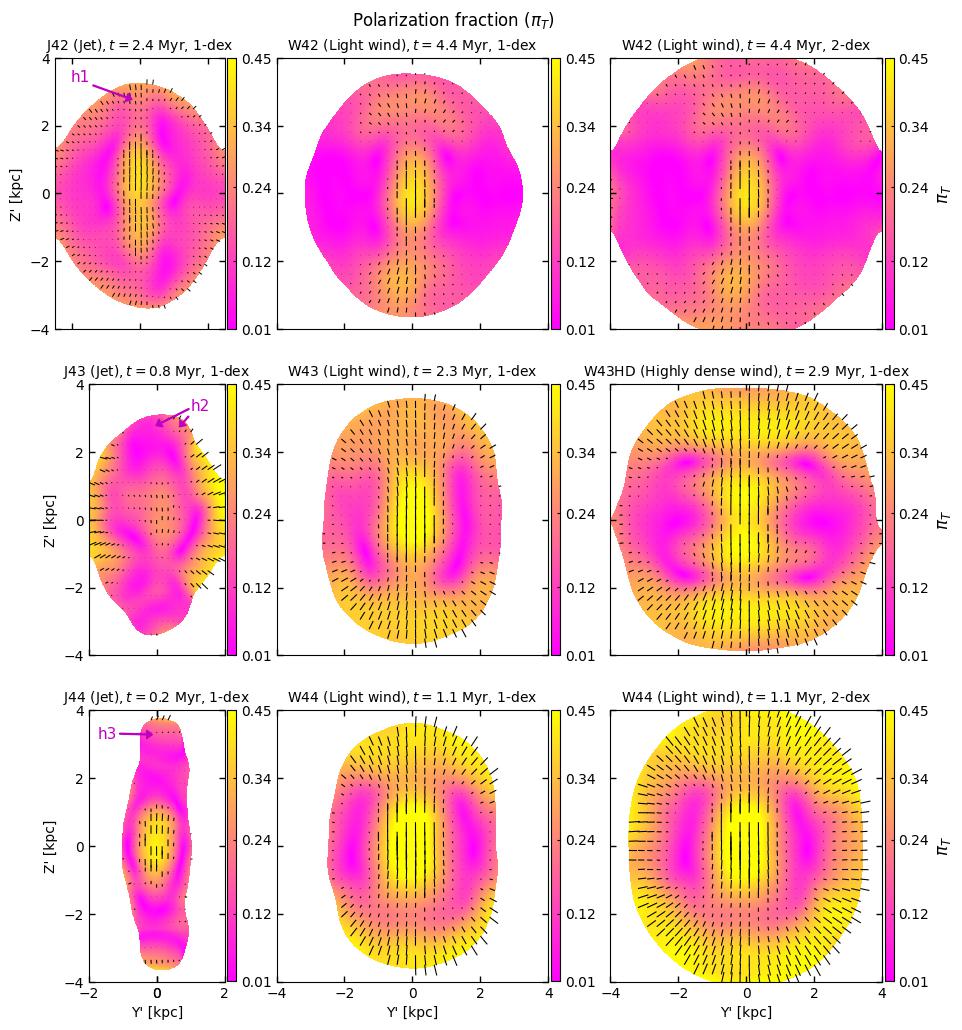} 
    \end{tabular}}

       \caption{Convolved polarization fraction maps corresponding to the emission maps displayed in Fig.~\ref{fig:synch_jet_wind_300pc}. The Gaussian filter used for convolution has a FWHM of 720~pc.}
\label{fig:pol_jet_wind_300pc}
\end{figure*}

The maps for the logarithmic synchrotron flux and polarization for the jets from Paper I and the winds from this study, observed at a viewing angle of $\theta_I=60^\circ$ are presented in this section. Specifically, the maps correspond to low resolution, where we employ a Gaussian filter with a FWHM of 720~pc for convolution, and are shown in Figs.~\ref{fig:synch_jet_wind_300pc} and~\ref{fig:pol_jet_wind_300pc}. Our focus primarily lies on displaying flux levels within a one-order dynamic range across all figures. However, for comparative purposes, we also include maps for W42 and W44 spanning a two-orders-of-flux range, which are shown in the top and bottom right plots. Regions below the lower limit of the flux color bar are omitted from both emission and polarization images.

\bsp	
\label{lastpage}
\end{document}